\documentclass[twocolumns]{aa} 

\usepackage{amssymb}
\usepackage{natbib}
\usepackage{graphicx,color}
\usepackage{txfonts}

\newcommand{\Ms}{\ensuremath{M_\odot}}

\newcommand{\el}[2]{$\rm{}^{#2}\kern-0.6pt#1$}
\newcommand{\elm}[2]{\rm{}^{#2}\kern-0.8pt\rm{#1}}

\begin{document}

\sloppy
  \title{Evolution of long-lived globular cluster stars}
   \subtitle{I. Grid of stellar models with helium enhancement at [Fe/H] = -1.75} 

\authorrunning{W. Chantereau et al.} \titlerunning{I. Evolution of enhanced helium stars}   

   \author{W. Chantereau \inst{1}
     \fnmsep\thanks{E-mail: william.chantereau@unige.ch}
          \and C. Charbonnel \inst{1,2}
          \and T. Decressin \inst{3} 
        }

   \institute{Department of Astronomy, University of Geneva, Chemin des Maillettes 51, CH-1290 Versoix, Switzerland
         \and
IRAP, UMR 5277 CNRS and Universit\'e de Toulouse, 14 Av. E. Belin, F-31400 Toulouse, France
         \and INAF – Osservatorio Astronomico di Roma, via di Frascati 33, I-00040 Monteporzio, Italy
   }
   
  \date{}
  
  \abstract
  {Our understanding of the formation and early evolution of globular clusters (GCs) has been totally overthrown with the discovery of the peculiar chemical properties of their long-lived host stars. 
  } 
  {As a consequence, the interpretation of the observed color-magnitude diagrams  and of the properties of the GC stellar populations requires the use of stellar models computed with relevant chemical compositions. }
  {We present a grid of 224 stellar evolution models for low-mass stars with initial masses between 0.3 and 1.0~M$_{\odot}$ and initial helium mass fraction between 0.248 and 0.8 computed for [Fe/H]=-1.75 
  with the stellar evolution code STAREVOL. This grid is made available to the community.
 }
  {We explore the implications of the assumed initial chemical distribution for the main properties of the stellar models: evolution paths in the Hertzsprung-Russel diagram (HRD), duration and characteristics of the main evolutionary phases, and the chemical nature of the white dwarf remnants. We also provide the ranges in initial stellar mass and helium content of the stars that populate the different regions of the HRD at the ages of 10 and 13.4~Gyr, which are typical for Galactic GCs.
  }
  {}


   \maketitle

\section{Multiple stellar populations in globular clusters:  Definitions and possible origins}
Globular clusters (GCs) 
are described in classical textbooks as  
old, massive, tightly bound systems hosting 
single stellar populations, i.e., 
coeval stars sharing the same initial chemical composition. 
These objects are extensively used as ideal benchmarks for stellar evolution theory, and their estimated absolute and relative ages provide 
markers for the formation and evolution of their host galaxies and their substructures \citep[e.g.,][]{Kruijssen2014}.

This idealistic picture has been shaken with the discovery of multiple 
stellar populations in 
individual Galactic GCs 
thanks to high-resolution  observations either with the Hubble Space Telescope or from the ground. 
Deep photometric studies have  revealed the presence of multiple sequences or spreads at the turn-off and/or along the subgiant branch in the color-magnitude diagrams of several GCs \citep[e.g.,][]{Bedin04,D'Antona05,Piotto07,Villanova07,Milone08,Milone10,Piotto08,Piotto09,Dalessandro11,King12,Milone13}. In addition, spectroscopic analysis is now commonly used to differentiate between so-called first- and second-generation GC stars (hereafter 1G and 2G) based on their positions along the well-documented O-Na anti-correlation
\citep[e.g.,][]{Prantzos06,Carretta09b,Carretta13}. 
This remarkable chemical pattern, which has been observed so far only among GC stars \citep[see, e.g.,][ and references therein]{Bragagliaetal14,Maclean15}\footnote{This feature has also recently been discovered among a small fraction of halo field stars (up to 2.5 \% of the star sample); see, e.g.,  \cite{Carretta10,Martell10,Ramirez12}.}, is advocated to be the main feature that distinguishes GCs from other stellar systems \citep[e.g.,][]{Carretta10}. 

Together with the C-N and Mg-Al anti-correlations, it is interpreted as the signature of hydrogen-burning at high temperature in massive, fast-evolving 1G GC stars (hereafter referred to as polluters) 
whose ejecta (depleted in carbon, oxygen, and magnesium, and enriched in nitrogen, sodium, aluminum, and helium) mixed to various degrees with original intra-cluster material and gave birth to 
2G stars  \citep[e.g.,][]{Prantzos07}. 

Different scenarios of secondary star formation invoke different types of 1G polluters, namely fast rotating massive stars (FRMS) with initial masses above 25 $M_{\odot}$ (\citealt{ Maeder06}; \citealt{ Prantzos06}; \citealt{Decressin07a,Decressin07b}; \citealt{Krause13}), massive asymptotic giant branch  (AGB) stars with initial masses between $\sim$ 6 and 11 $M_{\odot}$ (\citealt{Ventura01,Ventura13,D'ercole10,D'ercole11,D'ercole12,Ventura11}), and supermassive stars with initial masses around 10$^4$ $M_{\odot}$ (\citealt{DenissenkovHartwick14}).
In some cases the possible contribution of massive binary stars  \citep{DeMink09,Izzard13},  of FRMS paired with AGB stars \citep{Sills10},  
and of FRMS paired with high-mass interactive binaries \citep{Bastianetal2013a,CassisiSalaris2014} are also considered. 

Several studies have suggested that  the concept of stellar {\it generations} may not be appropriate to describe the long-lived stellar populations we are observing today in GCs, as this implies discontinuous star-forming events or long-term star formation. 
In particular the need for multiple star-forming episodes occurring 30 - 100~Myr after the formation of the 1G, as required by the AGB scenario, is in conflict with the star formation history derived for young massive clusters in the Milky Way and in Local Group galaxies \citep[e.g.,][]{Bastianetal2013b,CabreraZirietal2014}. 
An additional difficulty of this delayed star formation scenario lies in the need to re-accrete Galactic gas {\it \emph{with exactly the same chemical composition as that of the original proto-GC  material}} \citep[e.g.,][]{D'ercole11,D'ercole12} after 
several tens of Myrs and in any case after the expulsion of the original gas and of the SNe ejecta by SNe explosions or dark-remnant activation \citep{Krause13}. The probability that this has happened for all individual GCs during their Galactic journey is actually very small.

At the opposite end in the FRMS scenario, the formation of the sodium-rich, oxygen-poor low-mass stars (so-called 2G stars) is predicted to be completed before fast gas expulsion 
within $\sim$ 3.5 - 8~Myr after the formation of the polluters \citep{Krause13}. This timescale is compatible with the fact that very massive star clusters with ages of only a few Myrs in nearby galaxies appear to be devoid of gas \citep{BastianStrader14,Bastianetal14}. It is also shorter than the pre-main sequence evolution of the long-lived low-mass stars we observe today in GCs\footnote{The pre-main sequence lifetimes of our 0.8~M$_{\odot}$ models computed at the metallicity of NGC 6752 with initial helium mass fractions of 0.248, 0.4, and 0.8 
are $\sim 1.045 \times 10^8$, $4.32 \times 10^7$, and $2.87 \times10^6$ 
yrs, respectively (see Sect.~\ref{ImpactHelium0p8M}).}; low-mass stars formed out of original gas and those formed out of polluted material in the immediate vicinity of the FRMS should therefore have arrived on the zero age main sequence with very small time delay compared to the typical age of a GC of $\sim 13.4$ Gyr. It was also proposed that the ejecta of FRMS and/or high-mass interacting binaries were actually accreted by low-mass pre-main sequence stars of the same generation (\citealt{Bastianetal2013a}; see also Scenario II of \citealt{Prantzos06})\footnote{A possible difficulty is that in this case a very short accretion phase ($\sim$ 1 -- 2 Myr) is necessary  to account for helium enrichment and for the observed Li-Na anti-correlation \citep{D'Antona14,CassisiSalaris2014,SalarisCassisi14}.}, in which case the concept of distinct stellar generations is also misleading. 
Last but not least, in the FRMS polluter model a large percentage of (if not all) the stars exhibiting low sodium abundance (and that are therefore counted as 1G members; see \citealt{Prantzos06,Carretta09b}) could have formed from pristine material mixed with the H-burning ashes of the fast-evolving 1G polluters \citep{Charbonnel14}; 
in this case GCs would be formally devoid of 1G low-mass stars, their hosts being actually secondary star formation products with chemical properties spanning the whole range of the observed O-Na anti-correlations.
 
Consequently, to avoid conceptual as well as semantic ambiguities, we   adopt the denominations {\it ``first population"} (hereafter {\it 1P}) for the stars that were born with the original composition of the proto-GC, 
and {\it ``second population"} ({\it 2P}) for the stars that are born from a mixture between original material and the ejecta of short-lived {\it 1P} polluter stars.

\section{The importance of quantifying helium variations among GC stars}
\subsection{Dependence of the initial helium spread on the polluter scenario} 

Despite the many differences between the scenarios that have been developed to explain the abundance properties of GCs, it is widely accepted that the Na-enriched ({\it 2P}) GC stars 
have started their life with a higher helium content than their {\it 1P} counterparts (this is actually obvious as helium is the main product of hydrogen burning). 
This has been confirmed by direct spectroscopic measurements of non-local thermodynamic equilibrium He abundances for a subset of blue horizontal branch stars in one GC \citep[NGC 2808;][]{Marino14}.
However the {\it \emph{initial}} helium distribution among {\it 2P} stars is not well known yet, 
as predictions for the extent of helium enrichment along the sodium distribution in {\it 2P} stars strongly depend on the nature of the invoked {\it 1P} polluters.

In the case of the AGB scenario, the quantity of helium released by the polluters is set during the second dredge-up episode on the early AGB; it amounts to a maximum of $\sim$ 0.36 - 0.38 in mass fraction, independently of the initial mass of the AGB progenitor \citep[e.g.,][]{Doherty14}. 
On the other hand, the sodium content of the  ejecta of the polluter results from the competition between hot-bottom-burning and third dredge-up episodes that may occur during the later thermal-pulse AGB (TP-AGB) phase; therefore, it strongly depends on the initial mass of the polluter \citep{Forestini97,Siess07,D'ercole10,Ventura13}.
Consequently, {\it 2P} stars spanning a large range of Na abundances are all expected to  be born with very similar helium contents (maximum of 0.36 - 0.38 in mass fraction if no dilution with the ISM matter is taken into account  compared with $\sim$ 0.248 for {\it 1P} stars; see \S~\ref{comp1G}). 

In contrast, in the FRMS scenario the enrichment in sodium and helium of the polluters' ejecta results from central hydrogen-burning on the main sequence. 
Therefore this model predicts broad and correlated spreads in both sodium and helium among {\it 2P} stars at birth. In the case of  NGC~6752, for example, {\it 2P} stars are expected to be born with initial helium mass fractions ranging between 0.248 (as for the {\it 1P}) and 0.8 
\citep{Decressin07b}. 
Such a large helium dispersion correlated with the sodium spread provides a straightforward explanation to the lack of sodium-rich {\it 2P} AGB stars in this GC (\citealt{Charbonnel13}; observations by \citealt{Campbell13}) due to the impact of the initial helium content on the stellar lifetime and evolution path in the Hertzsprung-Russel diagram.
Simultaneously, the FRMS scenario predicts that {\it \emph{today}} (i.e., at a typical GC age of 13.4 Gyr), 95\% of the low-mass stars lying two magnitudes below the turn-off should be born with  a helium mass fraction Y$_{\rm ini}$ between 0.248 and 0.4, and only 5 \% with Y$_{\rm ini}$ higher than 0.4 \citep{Charbonnel13}. 
Additionally (see Sect.  \ref{mass_effect}), no star born with Y$_{\rm ini}$ higher than 0.4 is expected to lie today on the horizontal branch, in agreement with current interpretations of GC horizontal branch (HB) morphologies in connection with the so-called second parameter problem \citep[e.g.,][]{D'Antona02,D'Antona04,Caloi07,Gratton10,Dotter10}.

\subsection{Objectives of the present series of papers}

The aim of this series of papers is to investigate the implications of the initial helium spread predicted by the FRMS scenario in different parts of the GC color-magnitude diagram, and at various ages along GC evolution. 
Predictions will be compared to those of the AGB scenario and to available relevant observations, and future observational tests will be suggested.

Here (Paper I) we present a grid of low-mass, low-metallicity stellar models that will be extensively used for different purposes in the series. 
These models are computed with a [Fe/H] value of -1.75 close to that of the well-studied GC NGC 6752 and for a wide range of initial helium abundance, taking into account the corresponding variations expected within the FRMS model for the elements involved in the CNO cycle and the NeNa and MgAl chains (see \S~\ref{Sect:ModelInput} for a detailed description of the initial chemical compositions and  the input physics of the stellar models).
We discuss the impact of the initial chemical composition on the characteristics, lifetimes, evolution behavior,  
and fate of GC low-mass stars; we first focus on the case of a 0.8~M$_{\odot}$ star (\S\ref{ImpactHelium0p8M}), and then present the dependence on the initial stellar mass (\S\ref{mass_effect}). 
Although the influence of initial helium abundance on the stellar properties has already been extensively described in the literature \citep[e.g.,][]{Iben68,IbenRood1969,Demarque71,Sweigart78,D'Antona05,Caloi07,Salaris05,Maeder09,Pietrinferni09,Sbordone11,Valcarce12,Cassisi13,Kippen13}, to our knowledge this is the first investigation of the impact of helium content higher than $\sim$ 0.42 on the whole evolution (from the pre-main sequence to the end of the AGB phase) at low [Fe/H], and for such a broad mass range\footnote{\cite{D'Antona10} have investigated the effect of a $Y$ up to 0.8, but only on the horizontal branch morphology.}.

\section{Grid of {\it 1P} and {\it 2P} low-mass stellar models at [Fe/H]~=~-1.75}
\label{Sect:ModelInput}

\subsection{Generalities}

We present a grid of standard stellar models (i.e., with no atomic diffusion or rotation, and no overshooting) with initial masses between 0.3 and 1.0~M$_{\odot}$ (mass steps of 0.05~M$_{\odot}$) for [Fe/H] close to that of NGC~6752, and with Y$_{\rm ini}$ ranging between 0.248 and 0.8 and the abundances of the elements included in the CNO, NeNa, and MgAl chains that change accordingly (see \S~\ref{icc} for details on the initial chemical mixture assumed in the computations). The models are computed from the pre-main sequence up to the tip of the TP-AGB or the planetary nebula phase with the evolution code STAREVOL (e.g., \citealt{ Palacios06}). 
Mass loss is treated assuming the \cite{Reimers75} formula ($\eta$ parameter of 0.5) on the red giant branch (RGB) and during central helium-burning, and the   \cite{Vassiliadis93} prescription on the AGB.
The basic input physics (nuclear reaction rates, equation of state, opacities, mass-loss prescriptions) is the same as in \cite{ Lagarde12}, except for the mixing length parameter (1.75 here, as assumed in the FRMS models of \citealt{Decressin07b}, instead of 1.6).

\subsection{Initial chemical composition}\label{icc}
\subsubsection{{\it 1P} models}
\label{comp1G}

The initial composition of {\it 1P} stellar models must reflect that of the original intra-cluster gas. 
We assume an initial helium mass fraction Y$_{\rm ini}$ of 0.248 
(i.e., slightly higher than the primordial helium abundance 0.2463$\pm$0.0003 predicted by standard BBN by \citealt{Coc13} using the new value of $\Omega_b$ determined by \citealt{ Planck13}).
In order to be consistent with \citet{ Decressin07b} we adopt a scaled-solar abundance composition \citep{Grevesse93}, a heavy element mass fraction Z of 5 $\times 10^{-4}$,  and an alpha-enhancement of [$\alpha$/Fe] = + 0.3. 
This corresponds to [Fe/H] = -1.75, close to the value of -1.56 derived for NGC 6752 \citep[e.g.,][]{Carretta07a,Carretta09c,Villanova09}.

\subsubsection{{\it 2P} models}
\label{comp2G}

Following \citet{Prantzos06} and \citet{Decressin07a,Decressin07b}, we assume that 
the initial chemical composition of {\it 2P} stars consists of a time-dependent mixture between a) the intra-cluster material of the original composition and b) the material processed to various degrees through central H-burning and progressively ejected by the FRMS during their mechanically-driven slow wind phase (i.e., along the main sequence and the $\Omega \Gamma$-limit phase). 
As FRMS release the first material of original GC composition at the very beginning of their evolution, and later material richer in H-burning ashes, one expects   {\it 2P} stars with Na and He contents similar to that of {\it 1P} stars to form first, and then {\it 2P} stars with various degrees of Na and He enrichment (see, e.g., \citealt{Charbonnel14}). 

\begin{figure*}[!ht]
   \centering
   \includegraphics[width=.44\textwidth]{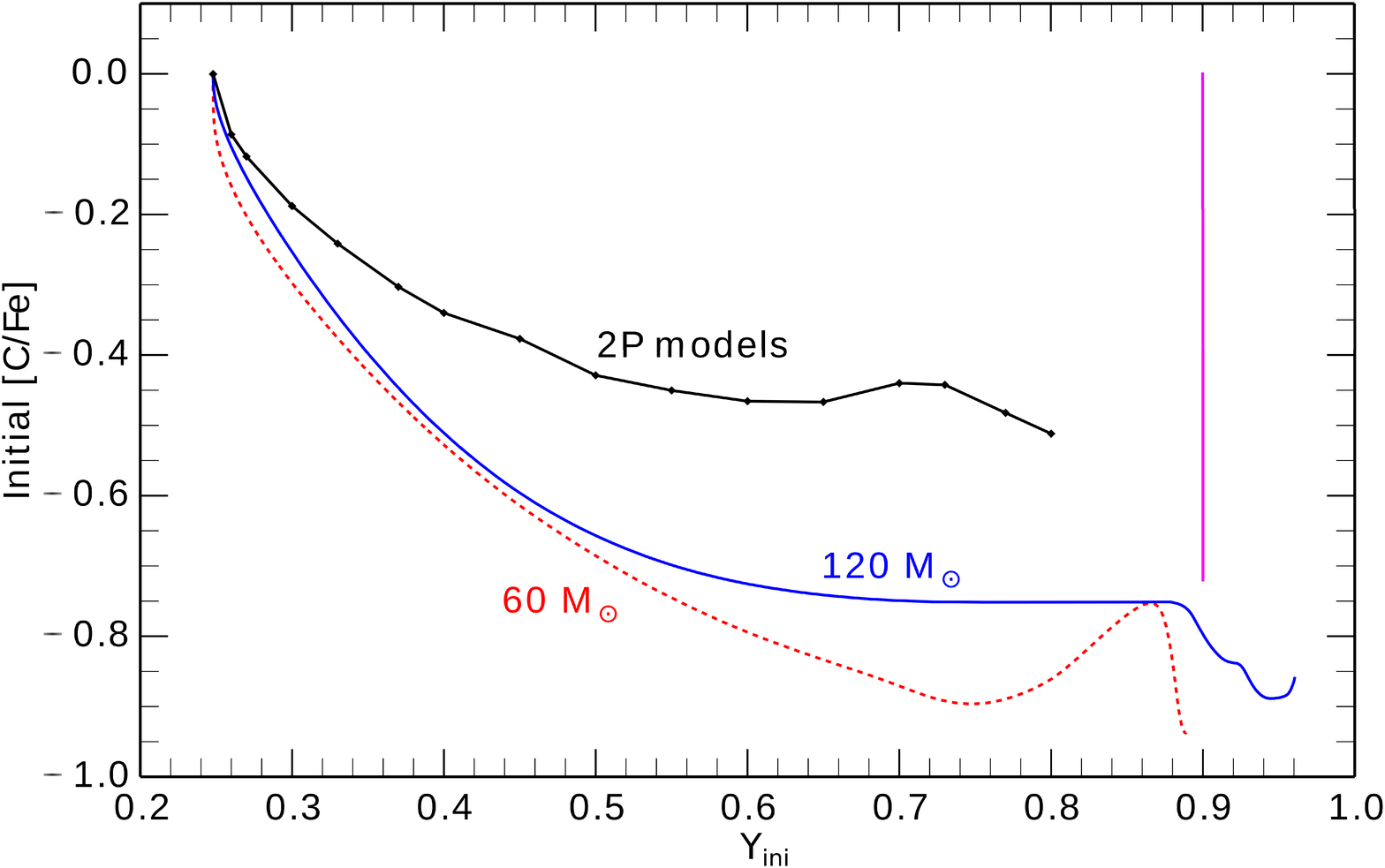} 
   \includegraphics[width=.44\textwidth]{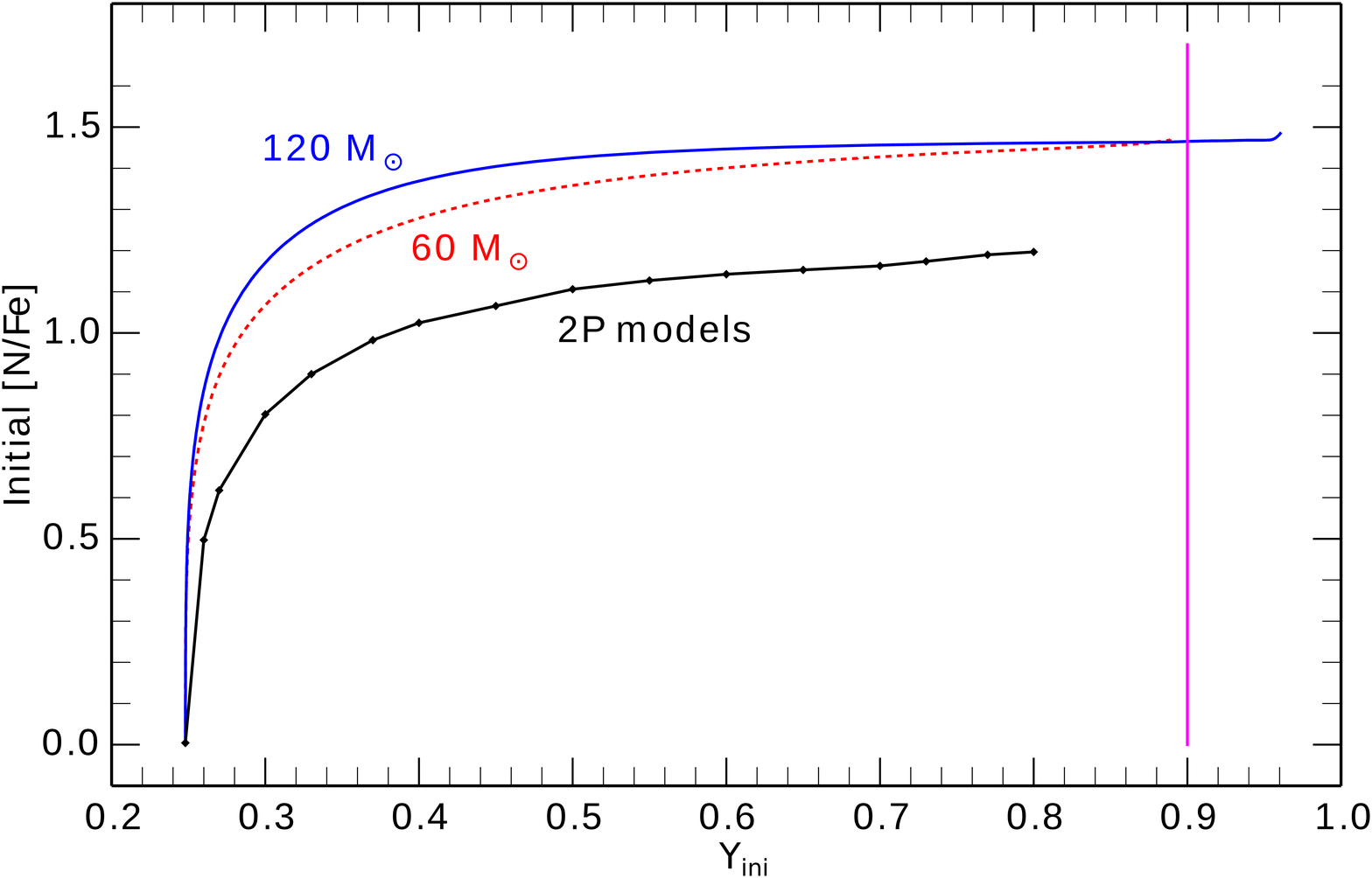}
   \includegraphics[width=.44\textwidth]{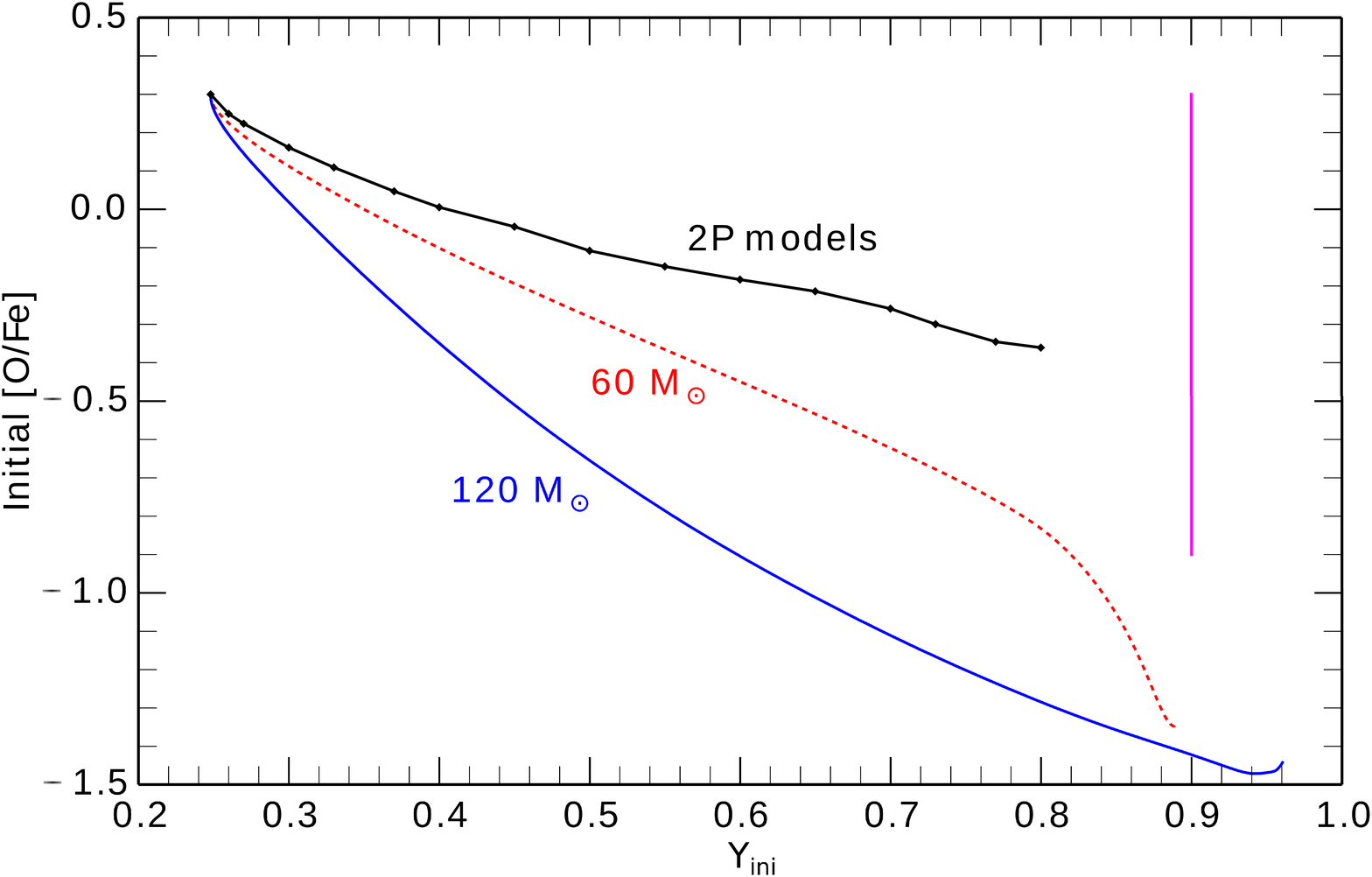}
   \includegraphics[width=.44\textwidth]{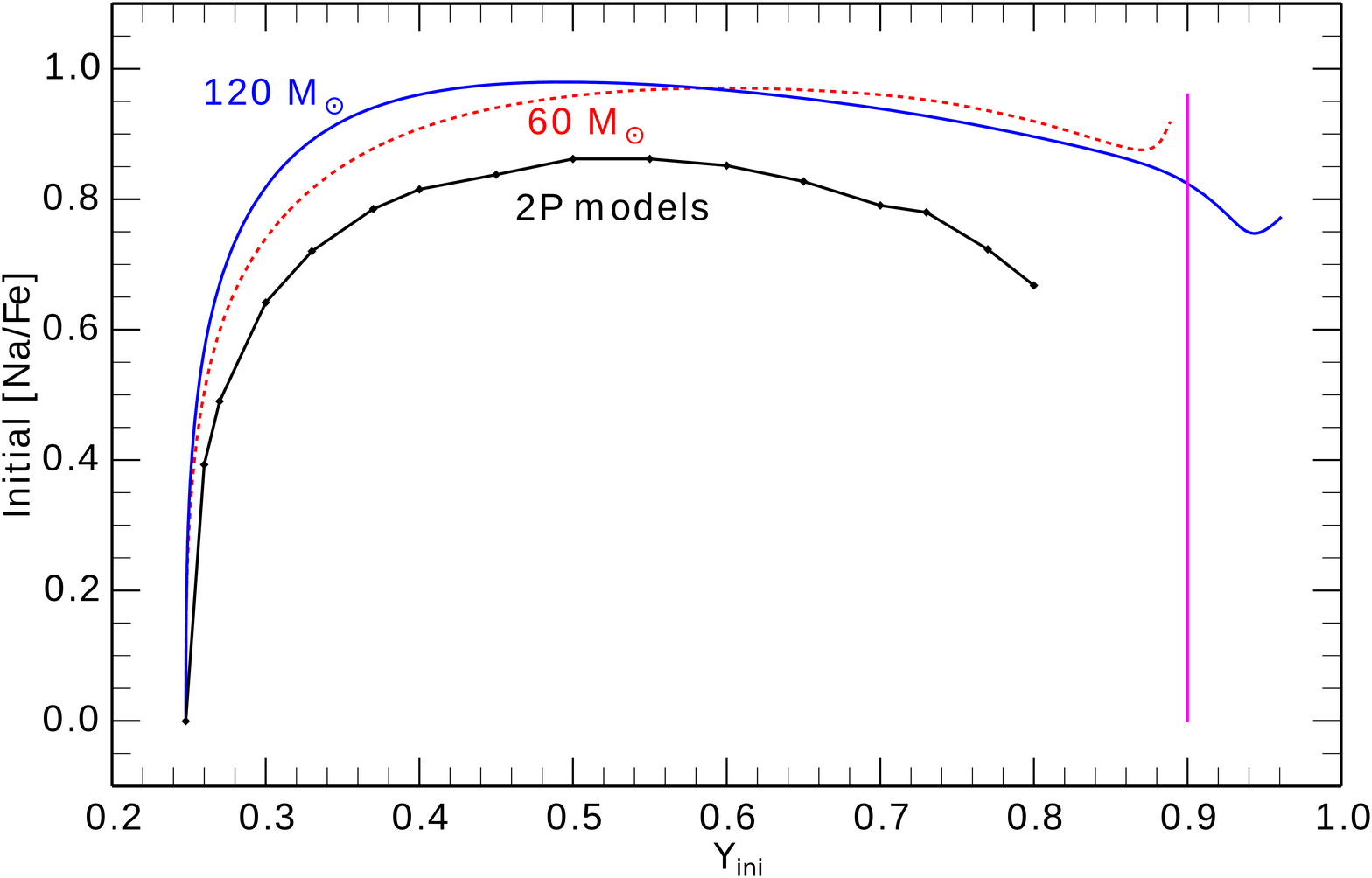}
   \includegraphics[width=.44\textwidth]{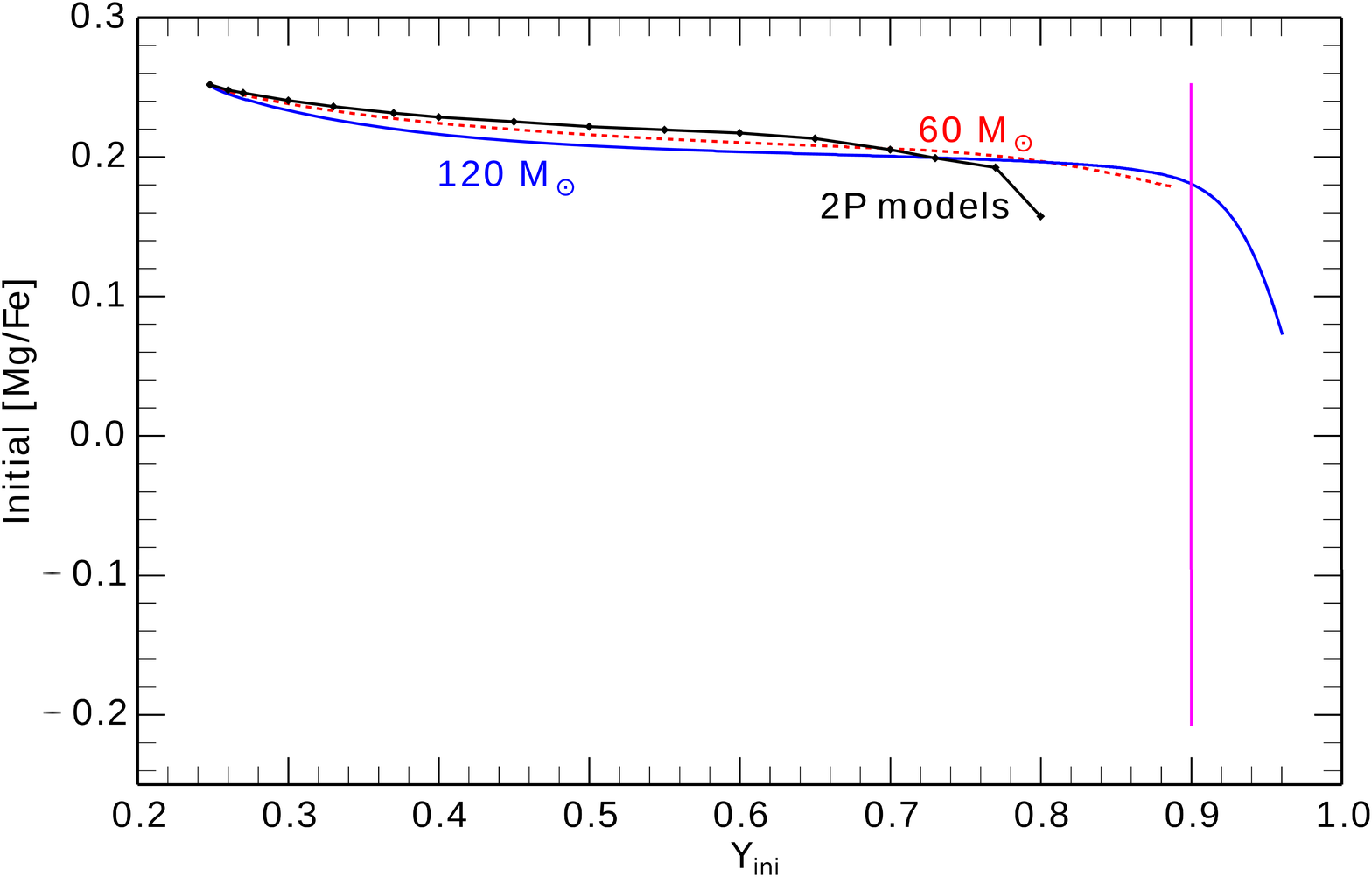}
   \includegraphics[width=.44\textwidth]{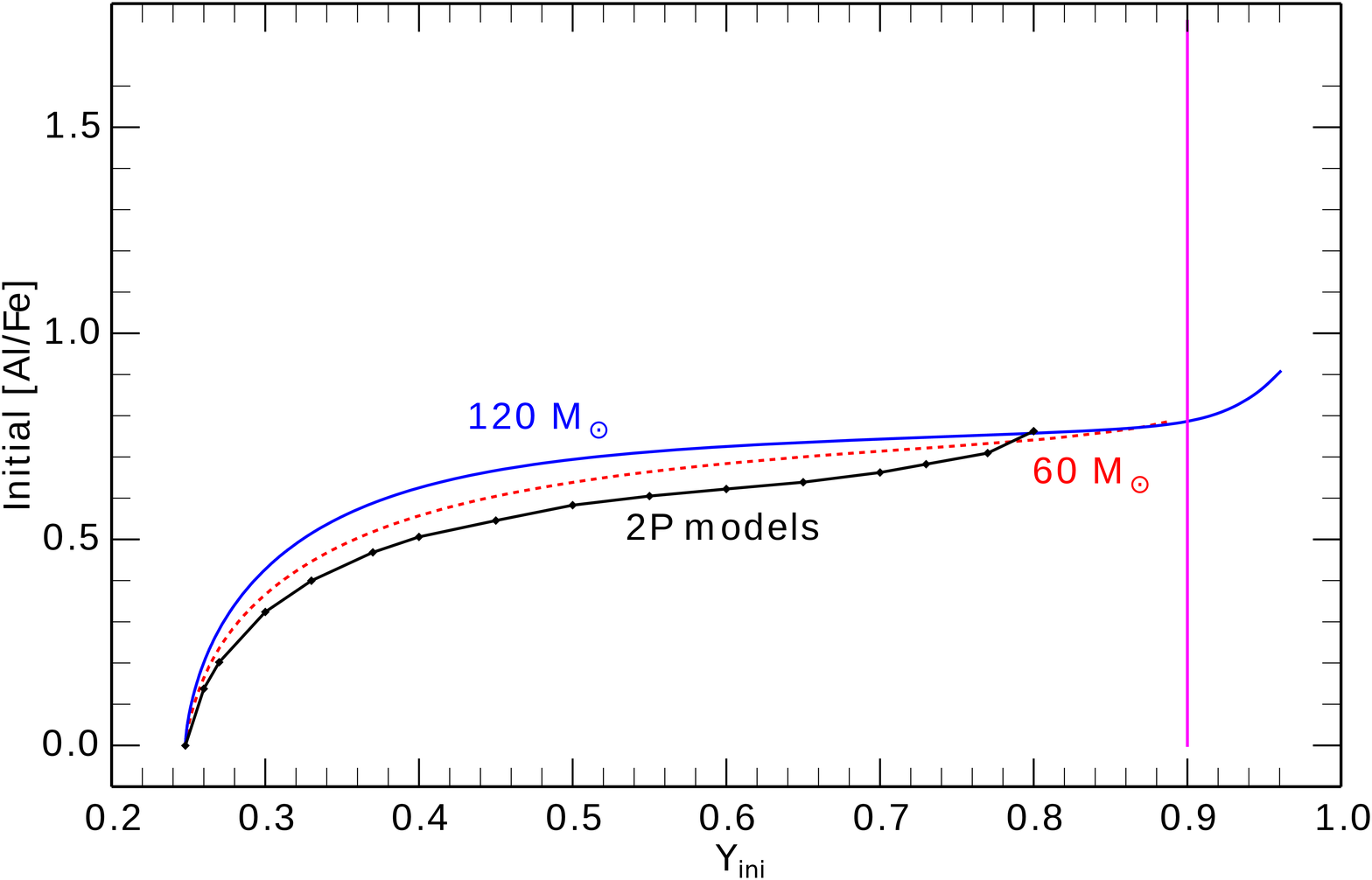}
   \includegraphics[width=.44\textwidth]{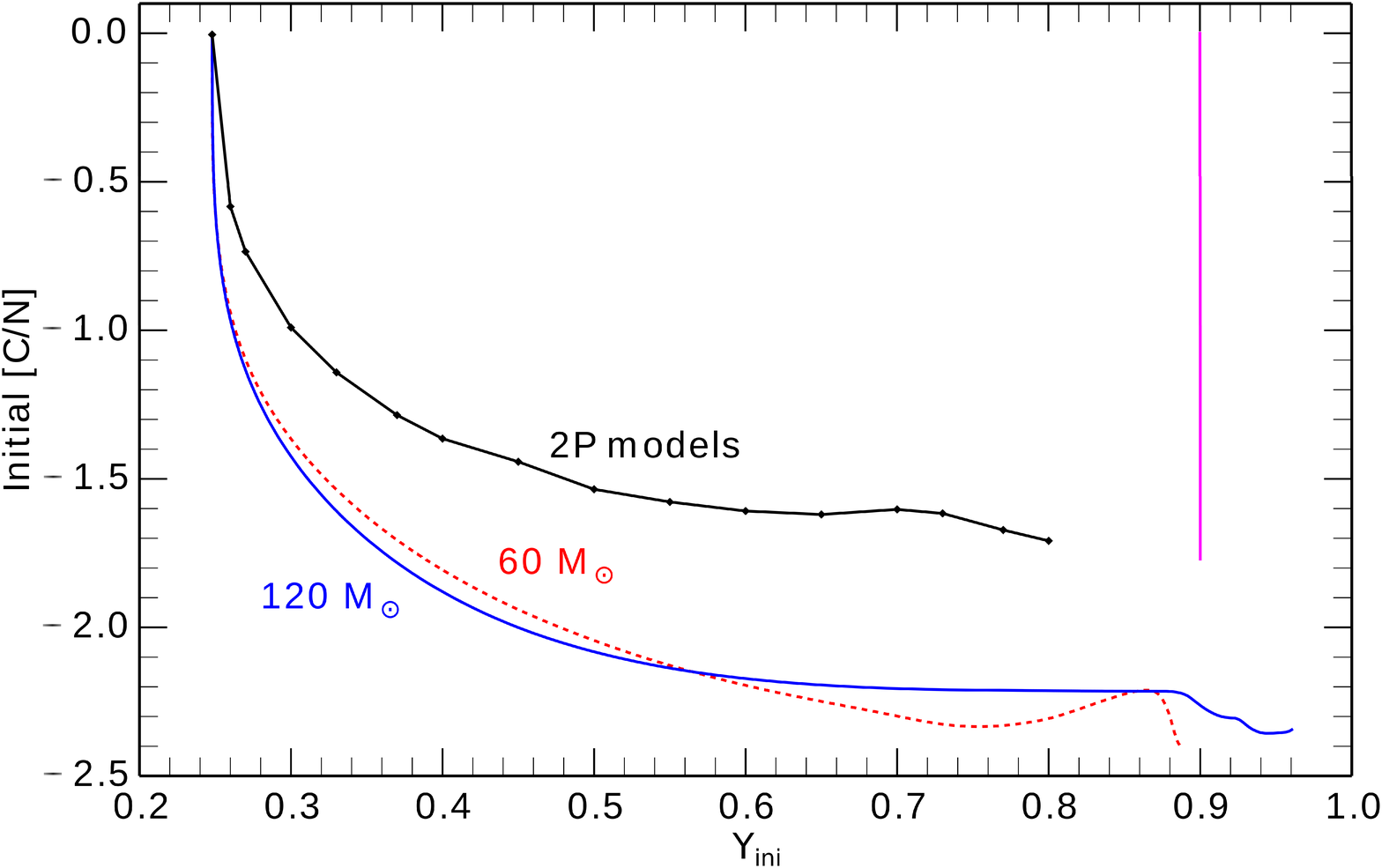}
   \includegraphics[width=.44\textwidth]{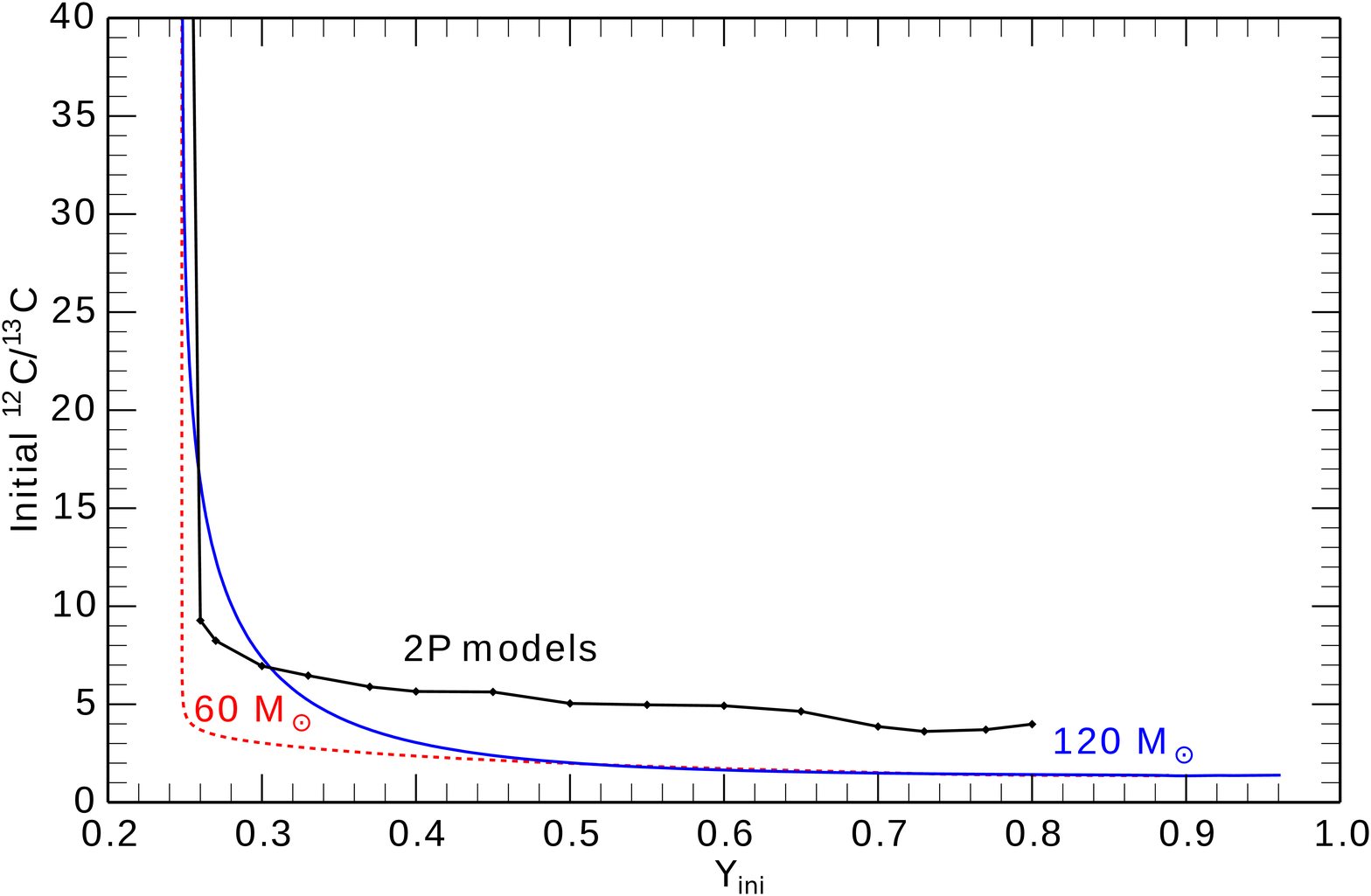}
    \caption{Theoretical distribution for the initial carbon, nitrogen, oxygen, sodium, magnesium, and aluminum abundances ( [X/Fe] = log(X/Fe)$_{\star}$-log(X/Fe)$_{\odot}$) for [C/N] and for $^{12}$C/$^{13}$C (initial value of 90) as a function of the initial helium mass fraction adopted for our {\it 2P} stars models (black line; dots indicate the 16 values adopted for the model computations of each stellar mass between 0.3 and 1~M$_{\odot}$). This accounts for dilution between original {\it 1P} material and the ejecta of the 60 and 120~M$_{\odot}$ FRMS models (red and blue lines, respectively; \citealt{ Decressin07b}) as described in Eq.~1 (see text).
The magenta arrows represent the variations observed in NGC 6752 for [C/Fe] and [N/Fe] \citep{Carretta05} ,  [O/Fe] and [Na/Fe] \citep{Carretta07a}, and [Mg/Fe] and [Al/Fe]  \citep{Carretta12}} 
  \label{Fig:AbondancesInitiales}
\end{figure*}

To account for this general behavior, we assume that the initial mass fraction of a given isotope $i$ in {\it 2P} stars
varies from star to star according to 
\begin{equation}
  \label{2Gabun}
  X^\text{2P}_{i} = X^\text{wind}_{i}~(1-a_\text{t})+X^\text{ICM}_{i}~a_\text{t},
\end{equation}
with the various terms described below. 
\begin{itemize}
\item The original intra-cluster medium value $X^\text{ICM}_{i}$ is equal to $X^\text{1P}_{i}$ assumed for  {\it 1P} stars (\S~\ref{comp1G}).
\item For the mass fraction of the isotope $i$ in the massive star ejecta, $X^\text{wind}_{i}$, we use the time-dependent predictions along the main sequence and the $\Omega \Gamma$-limit phase of the 60~M$_{\odot}$ 60rC model FRMS model of \citet{ Decressin07b} computed with the Set C of nuclear reactions (see their Table~2, and the discussion in \S~\ref{compobs}). In order to cover the highest helium enrichment, we also take into account the part of the ejecta of their 120~M$_{\odot}$ 120rC FRMS model that correspond to a helium mass fraction higher than 0.88 (i.e., the maximum value for the 60~M$_{\odot}$ star; see Fig.~\ref{Fig:AbondancesInitiales}). 
\item The dilution factor a$_\text{t}$ varies during the lifetime of individual {\it 1P} polluters between 0.99 (ejecta strongly diluted with the surrounding intra-cluster gas) and a minimum value a$_\text{min}$ of 0.2 that is constrained by the Na-Li anti-correlation observed in NGC 6752 turn-off stars (observations by \citealt{Pasquini05}; see, e.g., \citealt{Decressin07b}).
\end{itemize}

The resulting range for the initial abundances of C, N, O, Na, Mg, and Al of our {\it 2P} stellar models is shown in Fig.~\ref{Fig:AbondancesInitiales} as a function of the initial mass fraction of helium. 
For each stellar mass between 0.3 and 1.0~M$_{\odot}$ (mass steps of $\sim$0.05~M$_{\odot}$, i.e., 14 initial masses), we compute 16 models with initial helium abundance varying between 0.248 and 0.8 (steps $\le$ 0.05, actual Y$_{\rm ini}$ values indicated by the black points in Fig.~\ref{Fig:AbondancesInitiales}). 
Therefore, our grid consists of 224 stellar models in total. 

\begin{figure}[!ht]
   \centering
   \includegraphics[width=.44\textwidth]{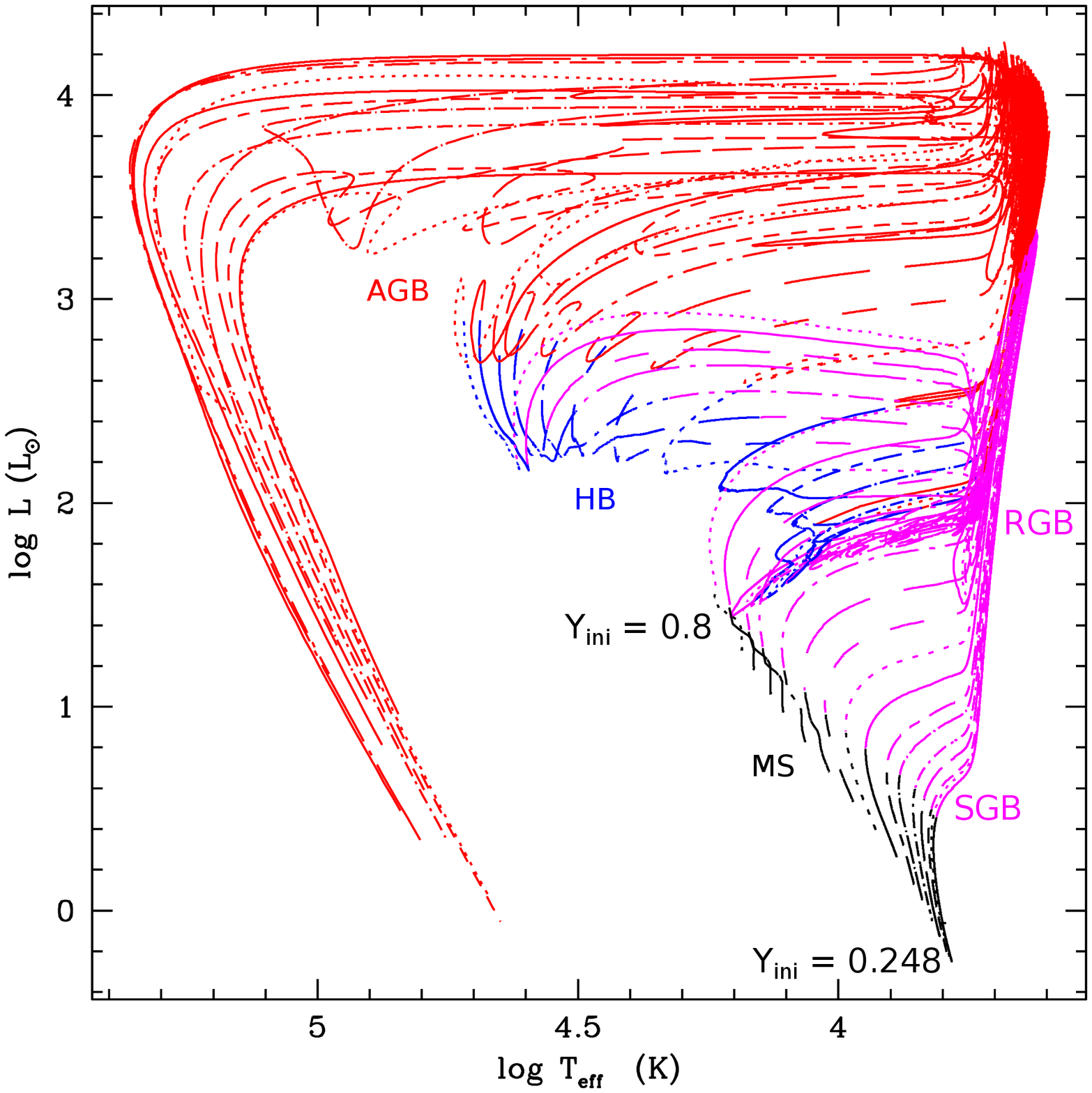}
   \includegraphics[width=.44\textwidth]{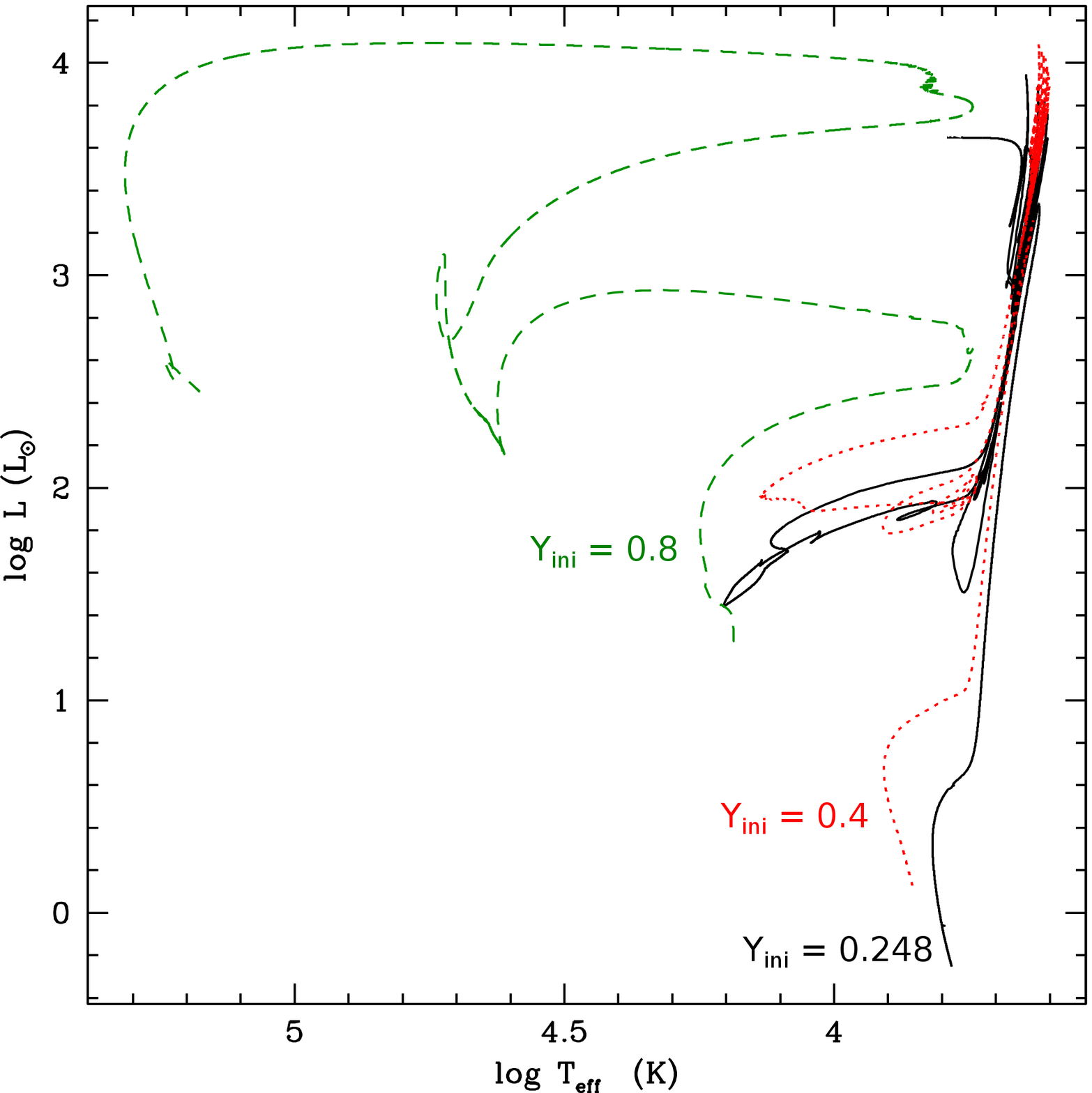}
    \caption{(Top) Evolution in the HRD of the 0.8 M$_{\odot}$ models for 16 different Y$_{\rm ini}$ between 0.248 and 0.8 
    (Z = $5 \times 10^{-4}$). Colors depict the evolutionary phases (main sequence in black, subgiant and red giant branch in magenta, central helium burning in blue, post central helium burning in red; the pre-main sequence is not shown for clarity).
    (Bottom) 0.8 M$_{\odot}$ models computed with Y$_{\rm ini}$ = 0.248, 0.4, and 0.8  (black, red, and green, respectively)}
  \label{hrd_0p8M_variousY}
\end{figure}

\subsubsection{Comparison with spectroscopic abundances in NGC~6752}
\label{compobs}
Within the FRMS scenario, {\it 2P} low-mass stars are expected to form in the immediate vicinity of the massive star polluters, which might actually all evolve in different environments (e.g., in terms of pristine gas available), depending  on their orbit \citep{Krause13}.
Individual FRMS polluters of similar initial masses are thus expected to give birth to  {\it 2P} of slightly different compositions as a result of the peculiar mixing between individual winds and pristine material (in other words, a$_\text{t}$ and a$_\text{min}$ might vary for individual polluters). 

Therefore, the abundance variations covered by our grid of  {\it 2P} stellar models represent the mean trends for all chemical elements, but do not cover the most extreme abundance variations that could be reached  in the case of  {\it 2P} stars formed out of raw FRMS ejecta, for example. This can be seen in Fig.~\ref{Fig:AbondancesInitiales}, where the initial chemical composition for our {\it 2P} models is compared with the maximum abundance variations that have been found among main sequence and red giant stars in NGC~6752 \citep[respectively 0.7, 1.7, 1.2, 1.0, 0.46, and 1.8 dex for C, N, O, Na, Mg, and Al;][]{Carretta05,Carretta07a,Carretta12}.
As discussed in  \citet{ Decressin07b}, the observed variations in Mg and Al require an increase in the $^{24}$Mg burning rate with respect to the experimental values (their 60~M$_{\odot}$ 60rD model), which is not accounted for in the  60~M$_{\odot}$ 60rC model used in \S~\ref{comp2G}.  However, this will not affect any of our findings since the Mg and Al content has no impact on the stellar evolution tracks or on the stellar lifetimes.

\section{Impact of extreme helium enrichment on stellar properties all along the evolution: The $0.8 M_{\odot}$ case}
\label{ImpactHelium0p8M}

As mentioned above, several studies of the influence of the initial helium content on stellar properties can be found in the literature, but only for relatively modest helium enrichment or for specific evolutionary phases.  
Here we recall the main trends and explore in detail the impact of very high initial helium content (i.e., mass fraction $~\geq~0.4$) all along the evolution.
We first describe here the 0.8~M$_{\odot}$ case, which  is the typical GC turn-off mass found in the literature\footnote{Our 0.8~M$_{\odot}$ {\it 1P} model with Y$_{\rm ini}$ = 0.248, Z = $5 \times 10^{-4}$ leaves the main sequence at an age of 13 Gyr.}, and later discuss the various evolution paths depending on the initial stellar mass (\S~\ref{mass_effect}). The limits of the various evolutionary phases are chosen as follows: 
\begin{itemize}
\item the ZAMS is defined either by a central mass fraction of hydrogen that has decreased by 0.6\% compared to its surface value, or  by a  central temperature that has reached $3 \times 10^7$K;
\item the turn-off is reached when the central mass fraction of hydrogen is lower than $10^{-7}$;
\item  the end of the subgiant branch (SGB, base of the RGB) is  the moment when the mass coordinate at the top of the H-burning shell (i.e.,  when $\le$ 10$~erg~g^{-1}~s^{-1}$) reaches a local minimum. 
\item the zero age horizontal branch (ZAHB) corresponds to the point where the helium mass fraction at the center of the star is 0.8.
\end{itemize}

Figure~\ref{hrd_0p8M_variousY} shows the evolution tracks of all the 0.8 M$_{\odot}$ models that we computed with the 16 different initial compositions described in \S~\ref{comp2G}; for clarity we also  show the tracks of three selected cases corresponding to Y$_{\rm ini}$=0.248, 0.4, and 0.8. 

\subsection{Main sequence}

\begin{figure}[ht]
   \centering
   \includegraphics[width=.47\textwidth]{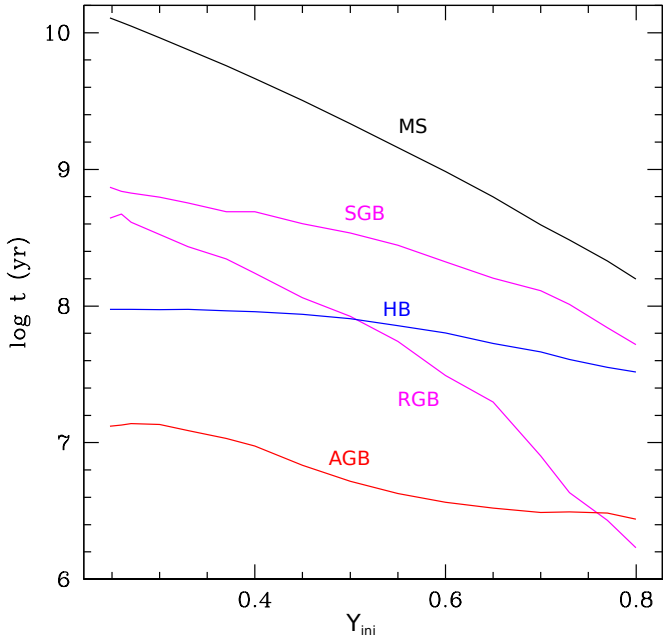}
    \caption{Duration of each evolutionary phase of the 0.8 M$_{\odot}$ models as a function of their initial helium mass fraction (main sequence, subgiant branch, red giant branch, horizontal branch, and asymptotic giant branch are shown in black, magenta, blue, and red, respectively)} 
  \label{lifetimes}
\end{figure}

When the initial helium content is higher, the opacity due to Thomson scattering is reduced in 
the stellar interior, 
and the nuclear burning luminosity increases during the main sequence because of the greater mean molecular weight\footnote{The homology relations give L $\propto \mu^{7.3}$ and  L $\propto \mu^{7.8}$ when the $pp$ chain or the CNO cycle dominates; see, e.g., \citet{Mowlavi98}.}.  Consequently, the evolution tracks are shifted towards higher luminosities and effective temperatures, and the main sequence lifetime is significantly shortened as shown in Fig.~\ref{lifetimes}: 
while the 0.8 M$_{\odot}$ model with Y$_{\rm ini}$=0.248 spends 12.9 Gyr on the main sequence (90 \% of its total lifetime), 
those with Y$_{\rm ini}$=0.4 and 0.8 have main sequence lifetimes of 4.62~Gyr and 158~Myr, respectively, which corresponds to 86 \% and 60 \% of their total evolution.

\begin{figure}[ht]
   \centering
   \includegraphics[width=.47\textwidth]{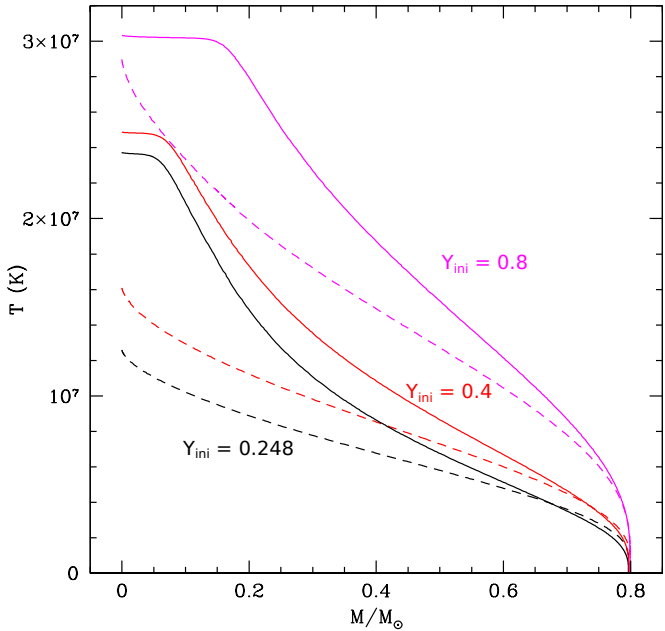}
    \caption{Temperature profiles at the zero age main sequence (dashed line) and at the turn-off (solid) inside the 0.8 M$_{\odot}$ models computed with the three different initial helium mass fractions}
  \label{temperatureprofile}
\end{figure}

\begin{figure}[!ht]
   \centering
   \includegraphics[width=.44\textwidth]{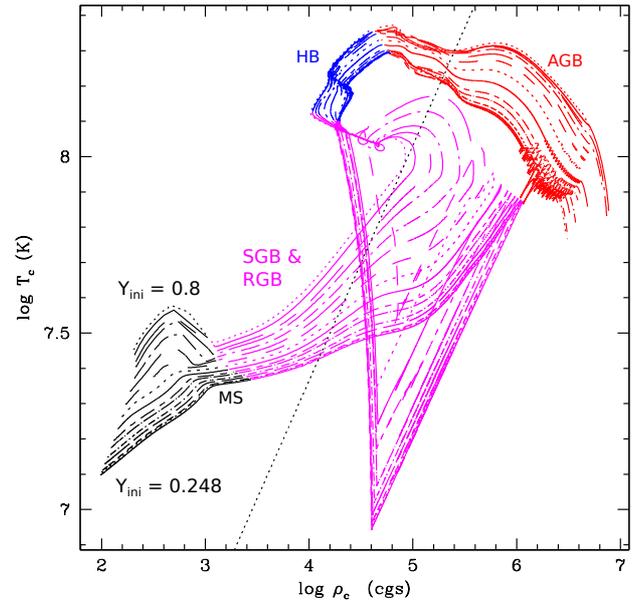}
    \caption{Central temperature as a function of the central density along the evolution of the 0.8~M$_{\odot}$ models computed with various initial helium mass fractions from 0.248 to 0.8 (colors correspond to evolutionary phases as in Fig.~\ref{hrd_0p8M_variousY}); the bend at the end of the RGB is characteristic of the helium flash.
    The straight dotted line corresponds to the limit 
    where the pressure given by the non-relativistic perfect gas law is equal to the pressure given by the completely degenerate non-relativistic gas, and where the mean molecular weight is   of the core of a RGB star, i.e., log(Tc) = 4.7 + 2/3 log($\rho_c$)}
  \label{logTclogRhoc}
\end{figure}

\begin{figure}[!ht]
   \centering
   \includegraphics[width=.4\textwidth, height=.35\textwidth]{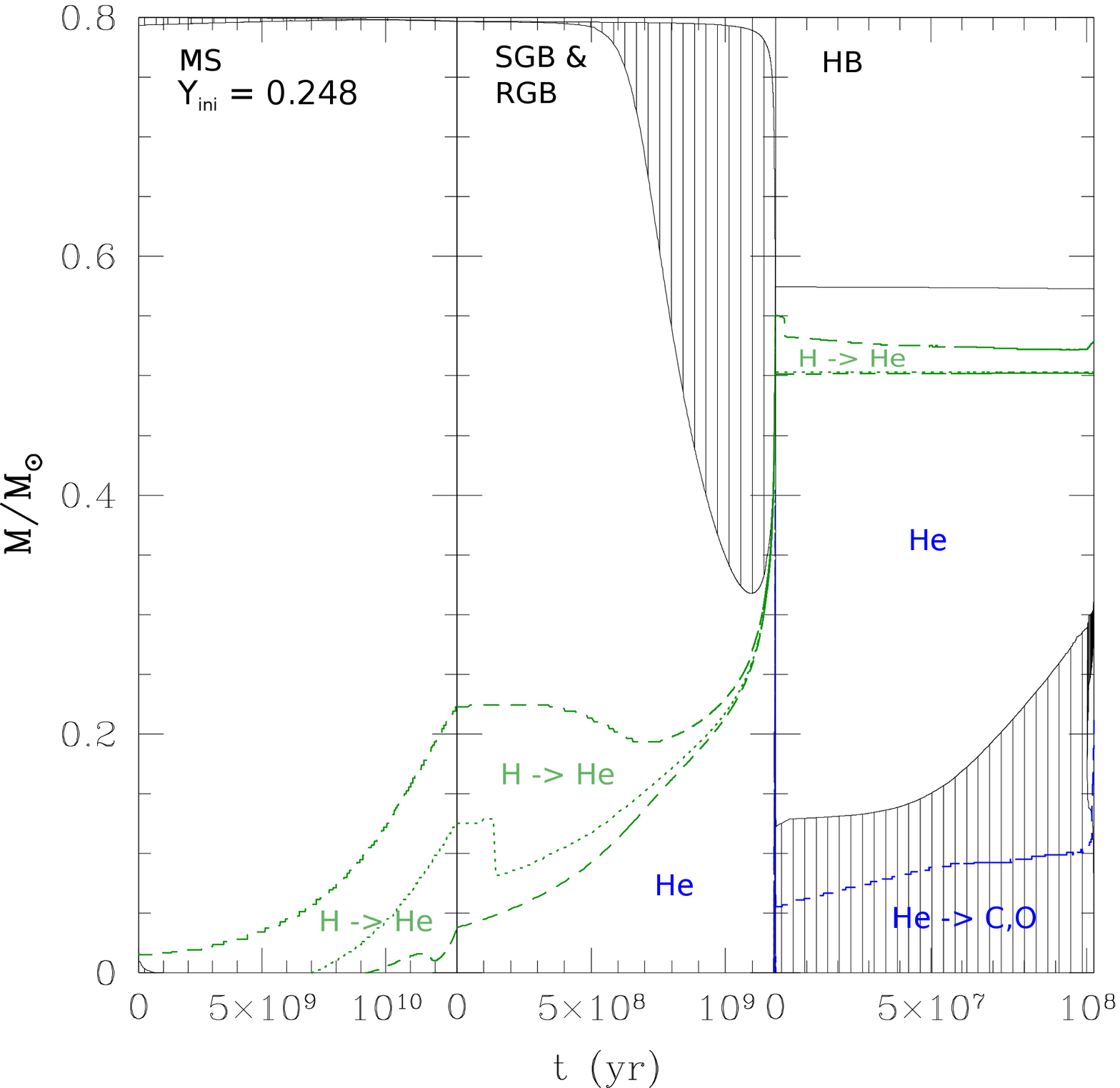}
   \includegraphics[width=.4\textwidth, height=.35\textwidth]{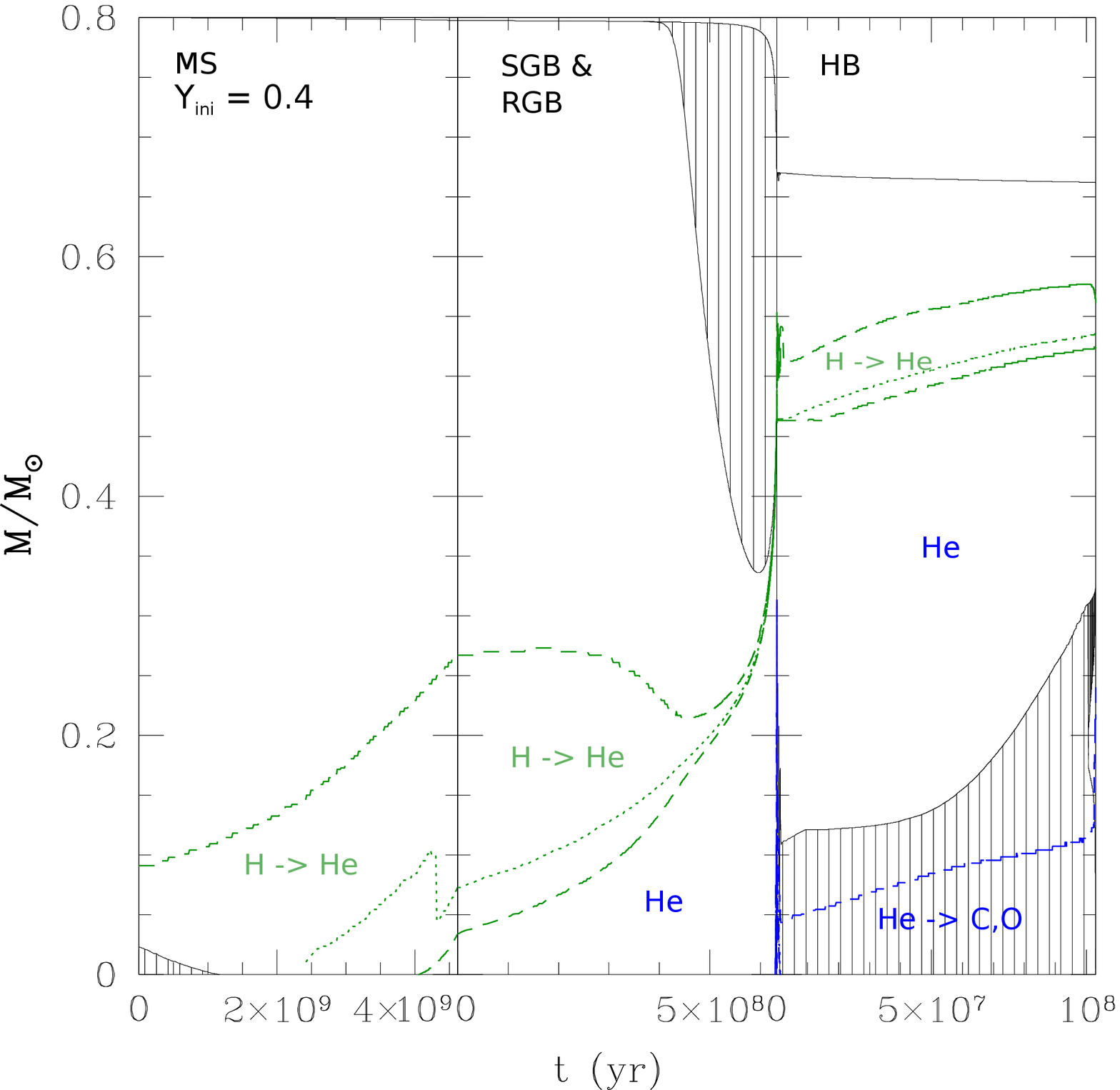}    
   \includegraphics[width=.4\textwidth, height=.35\textwidth]{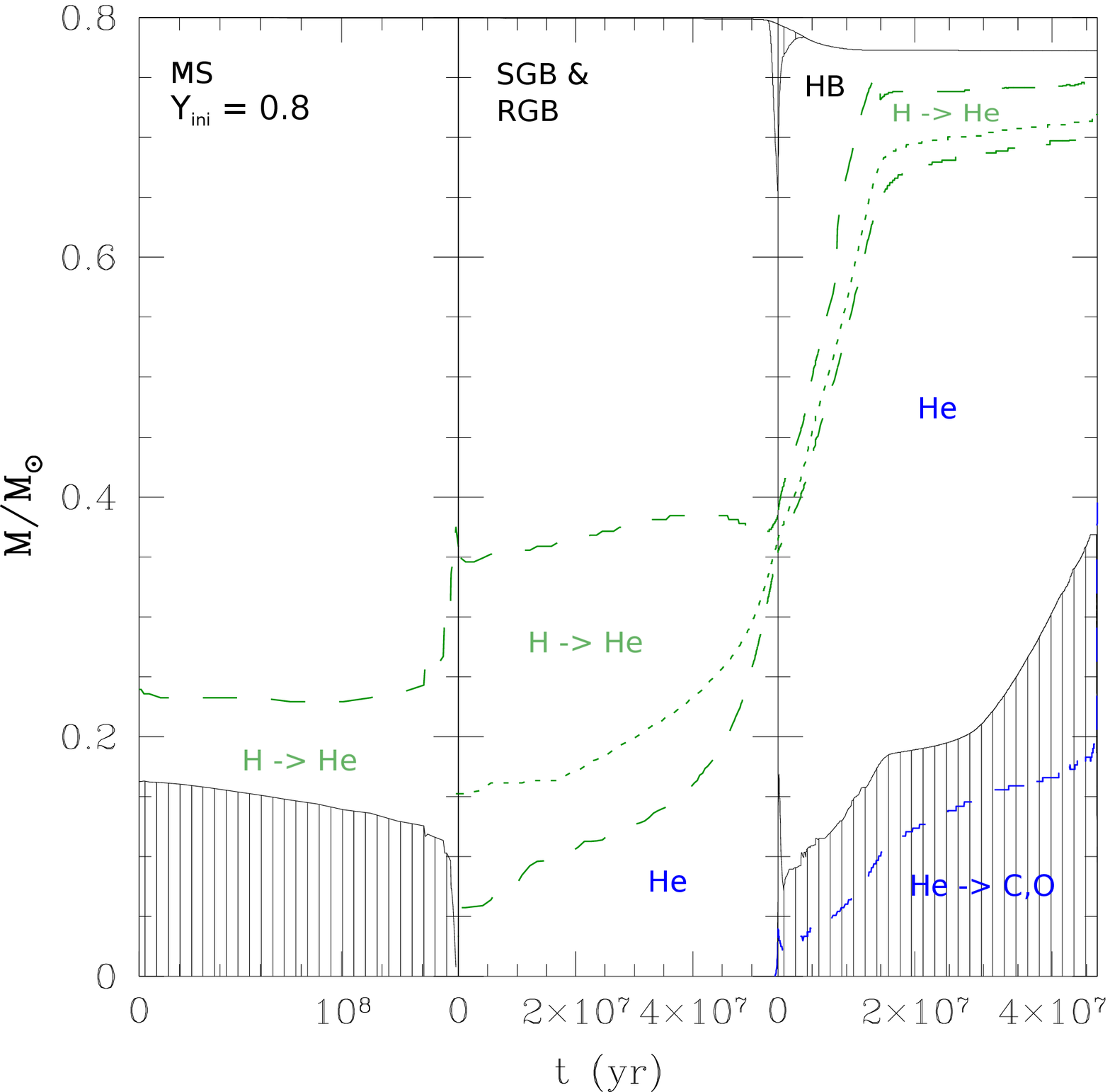}
   \caption{Internal properties (Kippenhahn diagram) of the 0.8 M$_{\odot}$ models with Y$_{\rm ini}$ = 0.248, 0.4, and 0.8 (top, middle, bottom) from the zero age main sequence to the end of central-helium burning. 
    Green dashed lines delimit hydrogen burning regions ($\ge$ 10$~erg~g^{-1}~s^{-1}$);
    blue dashed lines delimit the helium burning regions ($\ge$ 10$^{3}~erg~g^{-1}~s^{-1}$);  dotted green and blue lines show the maximum energy production of H- and He-burning; 
    hatched areas represent the convective zones. Timescales give the duration of each evolutionary phase}
  \label{kippenhahn}
\end{figure}

\begin{figure}[!ht]
   \centering
   \includegraphics[width=.4\textwidth, height=.35\textwidth]{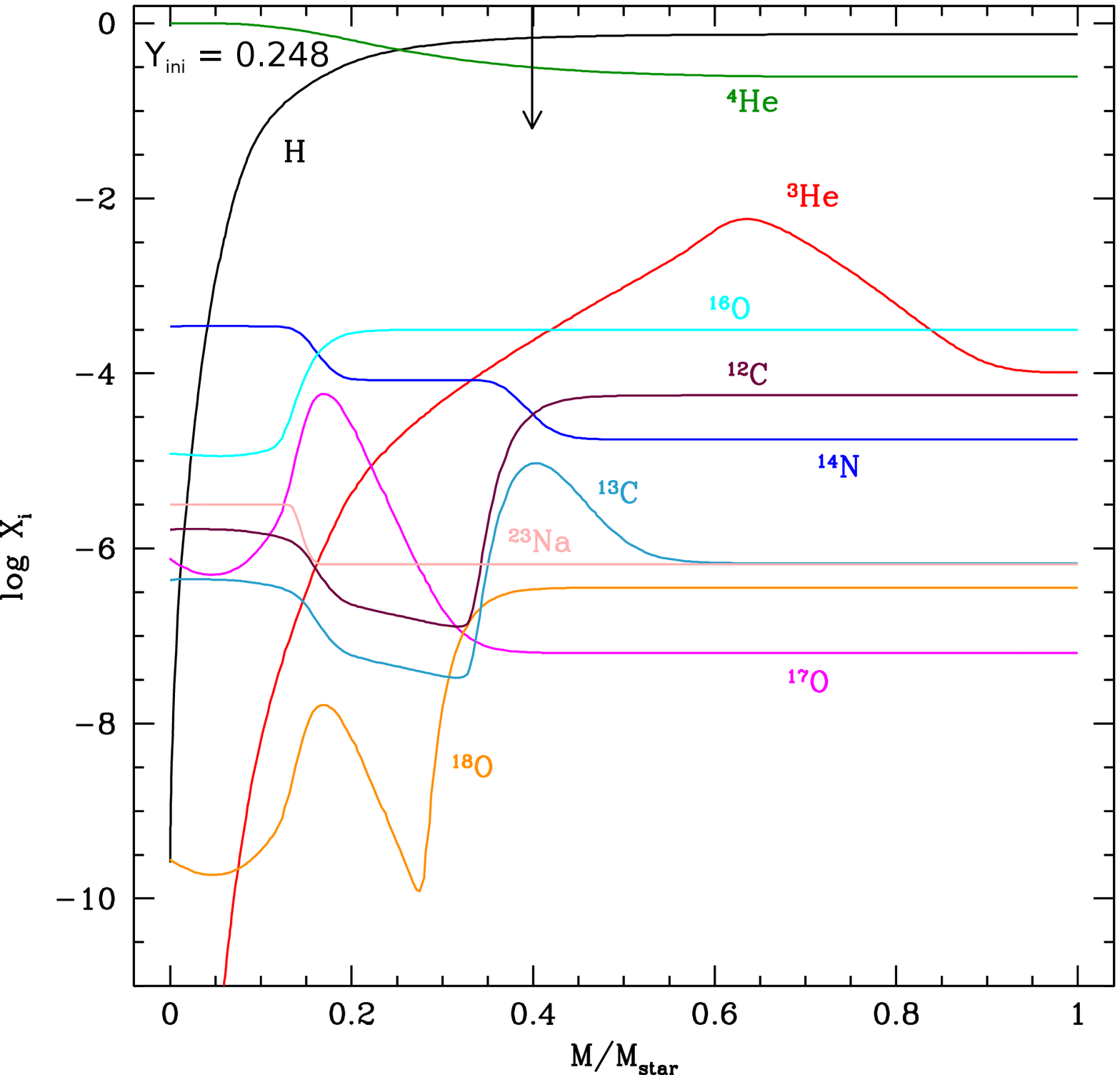}
   \includegraphics[width=.4\textwidth, height=.35\textwidth]{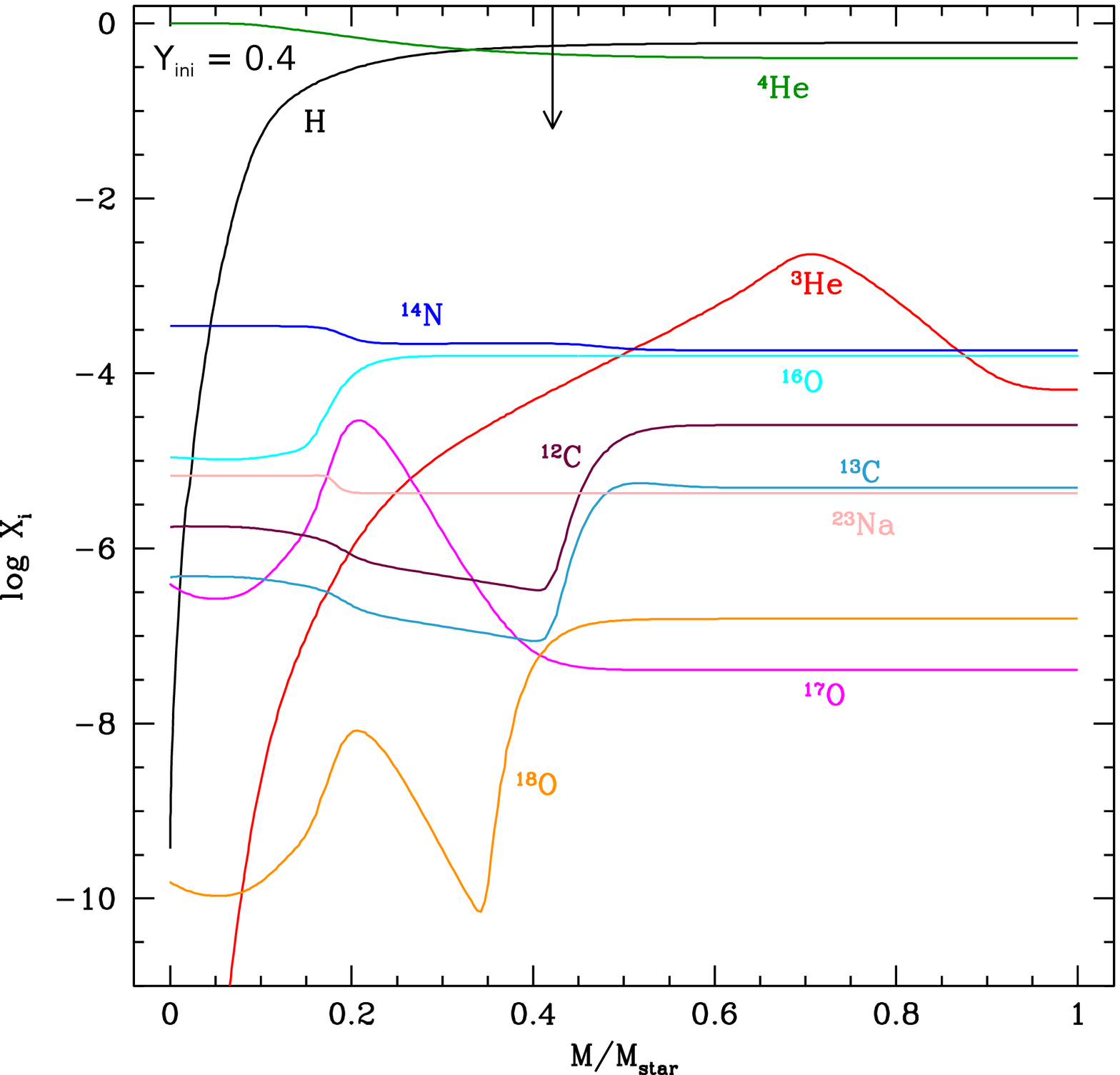}
   \includegraphics[width=.4\textwidth, height=.35\textwidth]{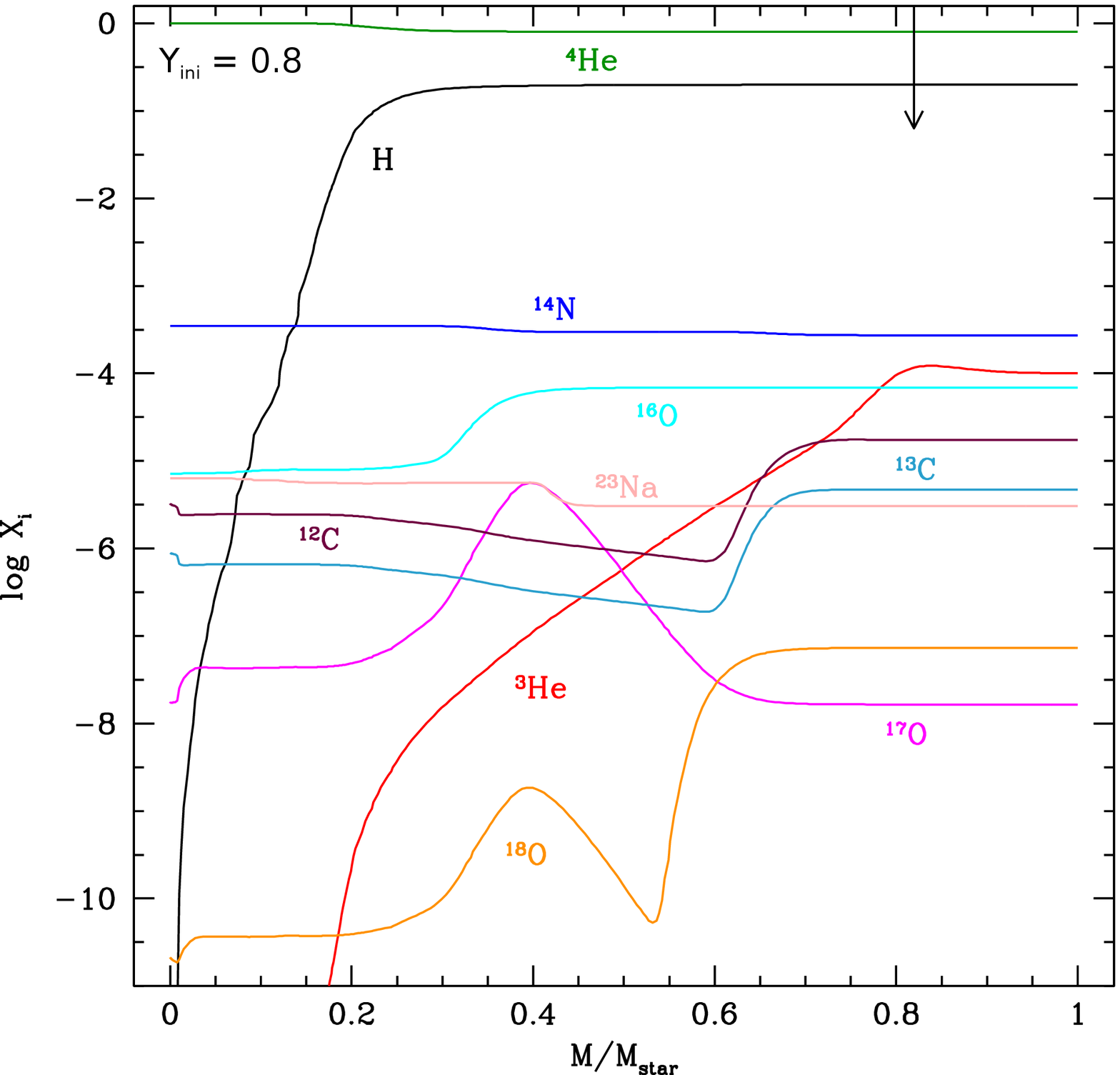}
    \caption{Abundance profiles of the main chemical elements as a function of depth in mass 
    at the end of the main sequence for the 0.8~M$_{\odot}$ models with Y$_{\rm ini}$=0.248, 0.4, and  0.8 (top, middle, bottom).
      The arrows indicate the maximum depth reached by the convective envelope during the first dredge-up on the red giant branch}
  \label{abundanceprofilesturnoff}
\end{figure}

During the main sequence, the internal temperature profiles steepen
and higher central temperatures are reached for higher initial helium content (Figs. \ref{temperatureprofile} and \ref{logTclogRhoc}).
This modifies the nuclear burning mode and the internal stellar configuration, as can be seen in the Kippenhahn diagrams presented in 
Fig.~\ref{kippenhahn}  for the three 0.8~M$_{\odot}$ models computed with Y$_{\rm ini}$=0.248, 0.4, and 0.8.
While the first two  burn central hydrogen mainly through the pp-chain and have a radiative core on the main sequence, the super He-rich model burns hydrogen only through the CNO cycle in a convective core (the transition occurs at Y$_{ini}$ = 0.5/0.55). In addition, when Y$_{\rm ini}$ is higher, the effects of central hydrogen burning are less pronounced because the initial abundances of all the isotopes involved in the CNO-cycle and the NeNa and MgAl chains are closer to their equilibrium values. We recall that the initial metal mixture at varying Y$_{\rm ini}$ is obtained from the dilution of matter coming from the H-burning regions of massive stars with pristine gas. Consequently, the internal chemical profiles of the various models strongly differ at the end of the main sequence (Fig.~\ref{abundanceprofilesturnoff}), which will have an impact during the advanced evolutionary phases.

\subsection{Subgiant and red giant branches}\label{SGB}

\begin{figure}[ht!]
   \centering
   \includegraphics[width=.47\textwidth]{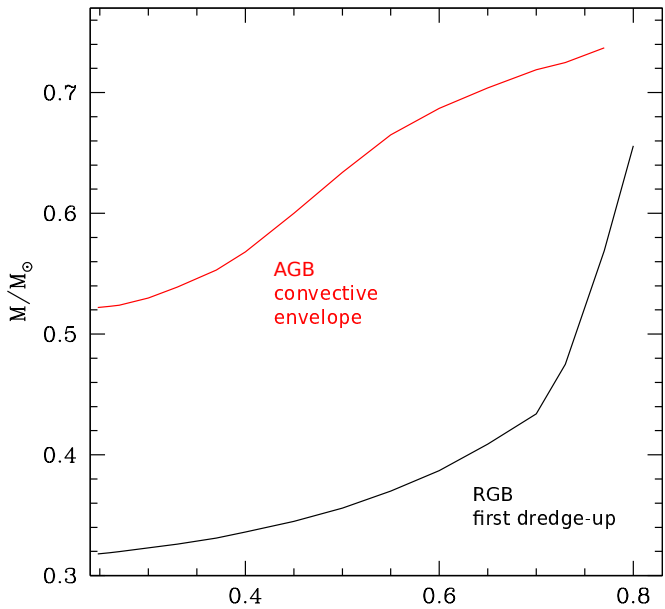}
    \caption{Mass layer (in solar mass) within the star reached by convection at its maximum extension on the RGB and AGB as a function of the Y$_{\rm ini}$ for the 0.8 M$_{\odot}$ models} 
  \label{envDUP}
\end{figure}

The models with higher initial helium content cross the Hertzsprung gap (subgiant branch) at higher luminosity. Although the first dredge-up occurs in all the 0.8~M$_{\odot}$ models of our grid, its effects on the surface abundances are smoothed out for higher Y$_{\rm ini}$ owing to shallower convective envelopes (see in Fig.~\ref{abundanceprofilesturnoff} the vertical arrows that indicate the maximum depth reached by the convective envelope during the first dredge-up; see also Figs.~\ref{kippenhahn} and \ref{envDUP}) and to initial chemical abundances closer to or already at equilibrium. This is quantified in Table~\ref{table_1}, which  summarizes the variations of the surface abundances during this phase for the three selected 0.8~M$_{\odot}$ models.  

\begin{figure}[!ht]
   \centering
   \includegraphics[width=.4\textwidth,height=.35\textwidth]{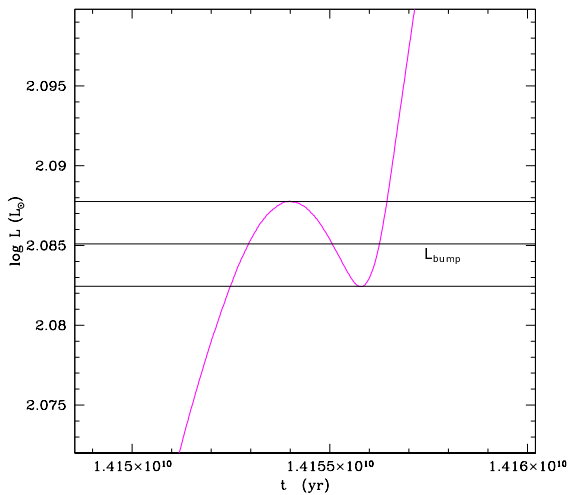}
    \caption{Variations of luminosity during the  RGB bump for the 0.8 M$_{\odot}$ model with Y$_{\rm ini}$ = 0.248. The luminosity starts dropping when the H-burning shell encounters the H-abundance discontinuity left behind by the convective envelope during the first dredge-up; it increases again when this discontinuity has been crossed. In Table~2 we give the time the star spends between the lower and upper horizontal lines, as well as the luminosity of the bump indicated by the middle horizontal line}
  \label{bump_parameters}
\end{figure}

The characteristics of the RGB bump are consequently affected.  While the 0.8~M$_{\odot}$ models with Y$_{\rm ini} \lesssim$ 0.4 undergo a drop in their total luminosity when the hydrogen-burning shell crosses the chemical discontinuity left inside the star by the convective envelope during the first dredge-up (Fig.~\ref{bump}), the super He-rich stars do not and have a constant increase of luminosity along the RGB evolution (see the characteristics of the bump in Table~\ref{tablebumpLrgbtip} and Fig.~\ref{bump_parameters}).

\begin{figure}[!ht]
   \centering
   \includegraphics[width=.47\textwidth]{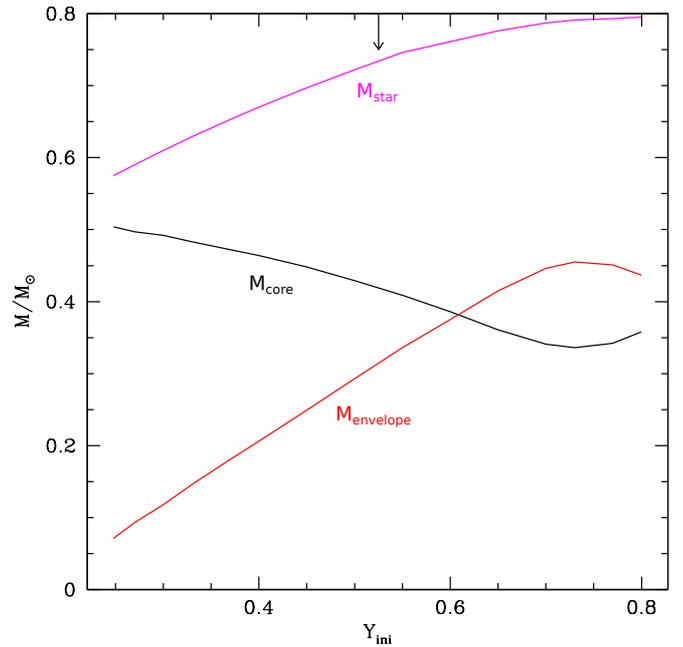}
    \caption{Mass of the helium core (black), of the convective envelope (red), and total stellar mass (magenta) at the RGB tip (i.e., at helium ignition) for the 0.8 M$_{\odot}$ models as a function of initial helium mass fraction. The arrow represents the maximum initial helium mass fraction for the He-flash to occur in the 0.8 M$_{\odot}$ case
    }
  \label{Mcoeurenvel}
\end{figure}

In the same way as for the other evolutionary phases, the lifetimes on the subgiant and the red giant branches shorten when Y$_{\rm ini}$ increases (Fig.~\ref{lifetimes}). 
This results from lower degeneracy and higher central temperature of the helium core (Fig.~\ref{logTclogRhoc}), which lead to smooth central helium ignition occurring at lower stellar luminosity 
for the super He-rich 0.8~M$_{\odot}$ models, while the models with Y$_{\rm ini}$ below $\sim$ 0.5/0.55 undergo the so-called helium flash at the RGB tip (see the different RGB paths in Figs.\ref{hrd_0p8M_variousY} and \ref{logTclogRhoc}).
Quantitatively, it takes 1.18 Gyr (8 \% of the total stellar lifetime) after the turn-off for the 0.8~M$_{\odot}$ model with Y$_{\rm ini}$ = 0.248 to ignite helium, 663 Myr (12 \%) when Y$_{\rm ini}$ = 0.4, and only 53.8 Myr (20 \%) if Y$_{\rm ini}$ = 0.8.
The amount of mass lost through the stellar wind during these phases (modeled here according to Reimers' prescription with $\eta$ = 0.5) decreases because of shorter timescales and lower luminosity ranging from 0.005 to 0.223 M$_{\odot}$. Therefore, the total stellar mass at helium ignition is lower for lower initial helium content (Fig.~\ref{Mcoeurenvel}).

\begin{figure}[!ht]
   \centering
   \includegraphics[width=.4\textwidth,height=.35\textwidth]{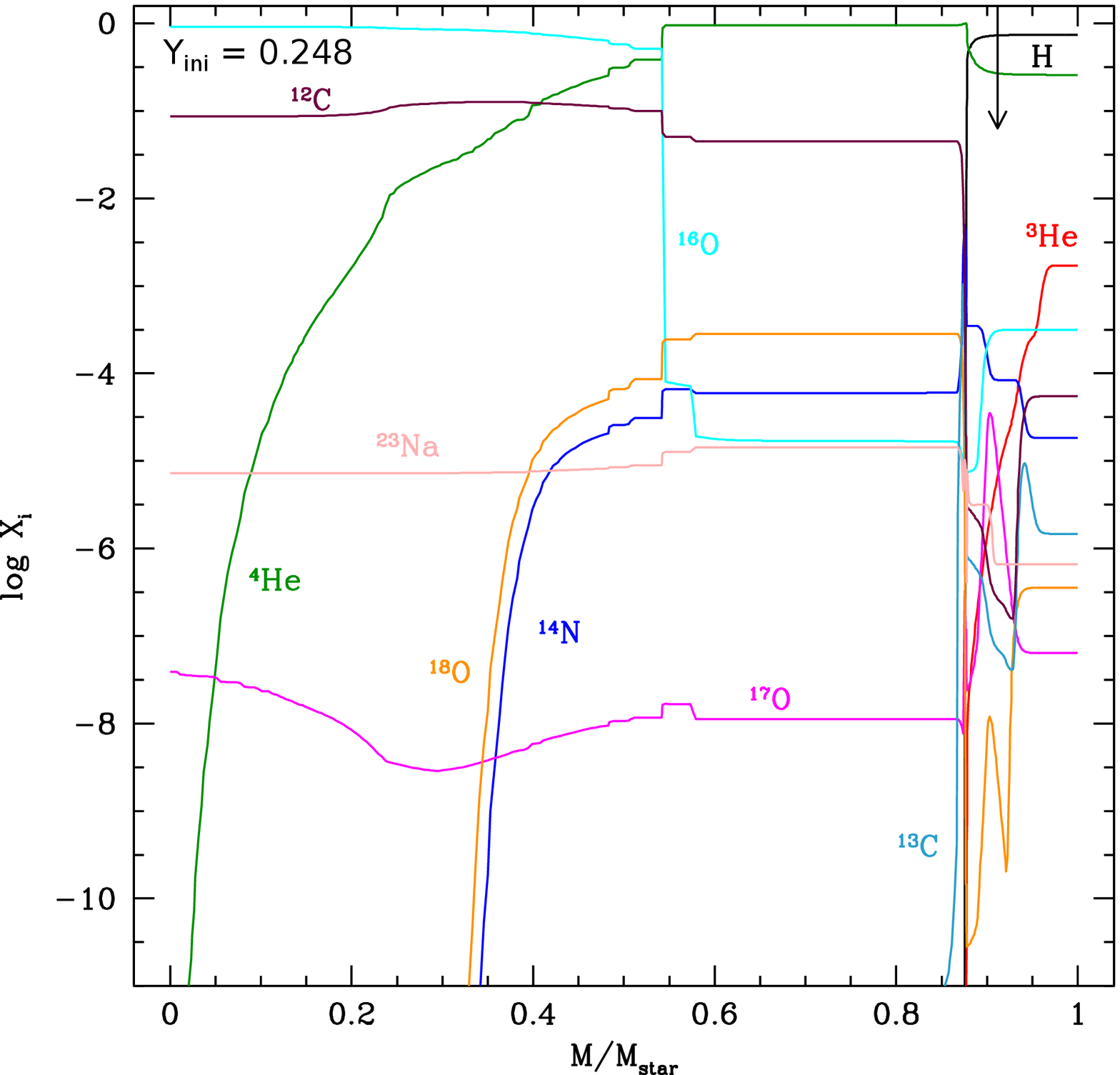}
   \includegraphics[width=.4\textwidth,height=.35\textwidth]{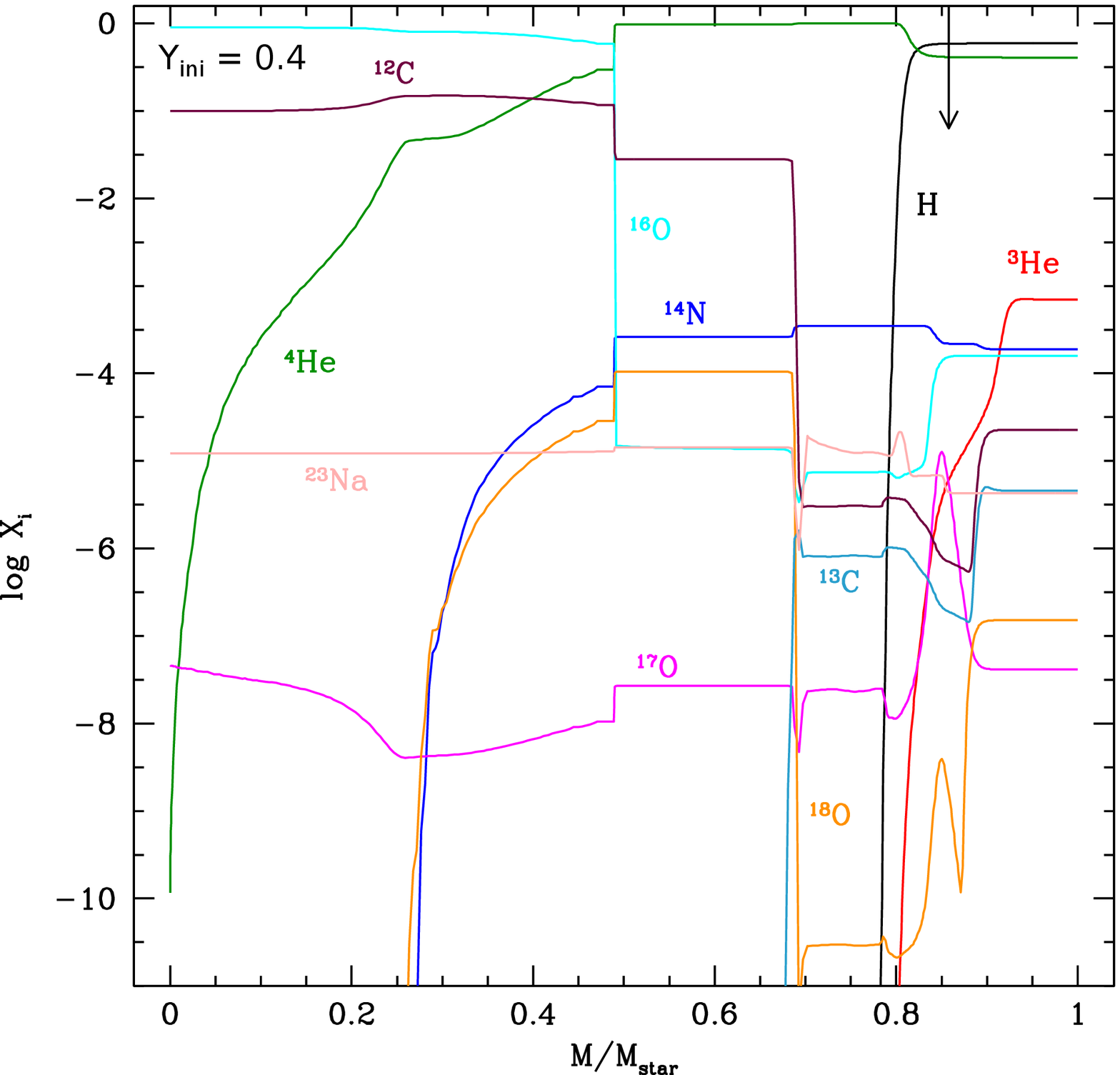}
   \includegraphics[width=.4\textwidth,height=.35\textwidth]{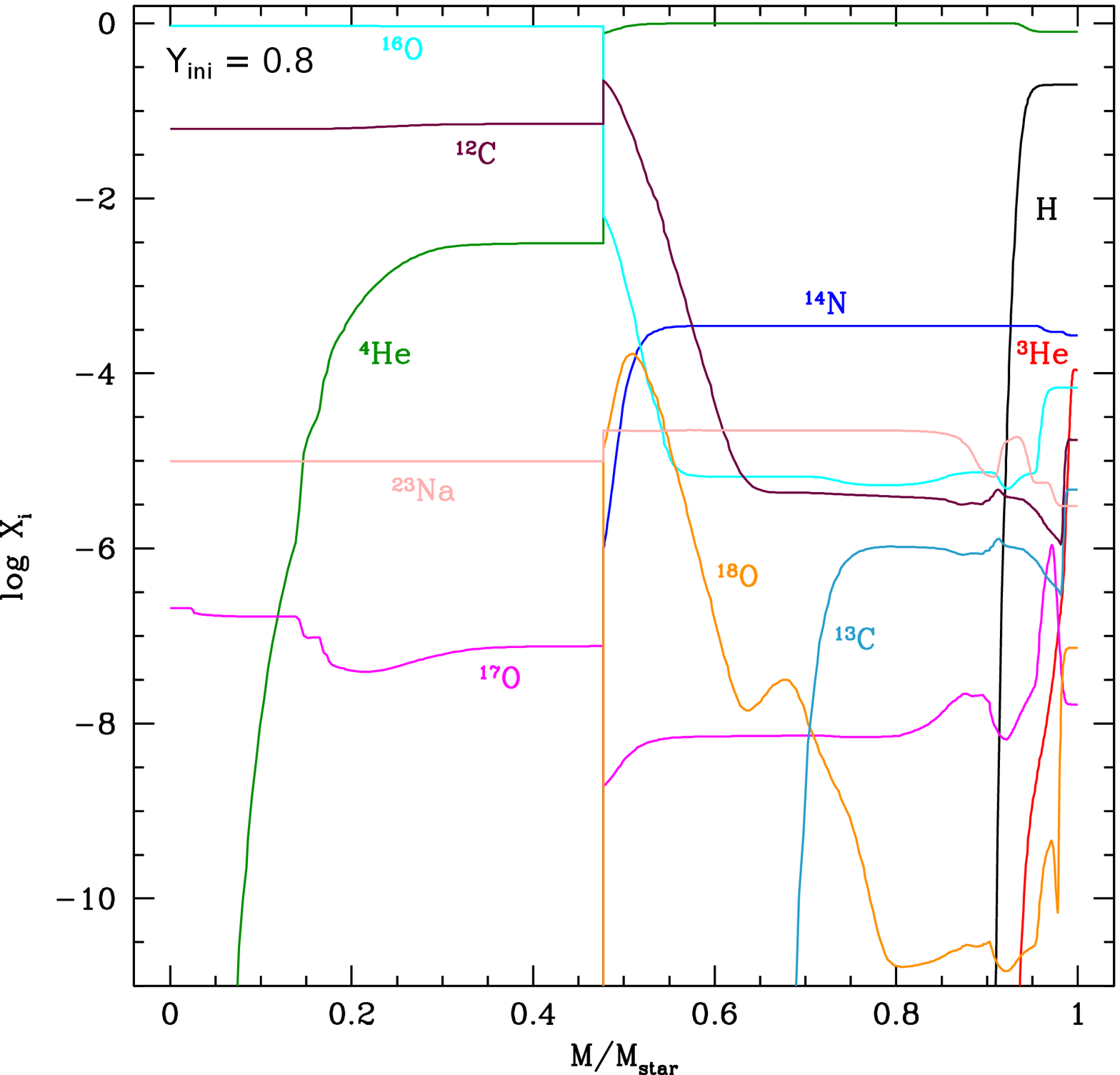}
    \caption{Abundance profiles as a function of the depth in mass for the main chemical elements at the end of the central helium burning for the 0.8~M$_{\odot}$ models with Y$_{\rm ini}$=0.248, 0.4, and 0.8 (from top to bottom).
    The total stellar mass at this evolution point is respectively 0.573, 0.662, and 0.772~M$_{\odot}$. 
    The arrows indicate the maximum depth of the convective envelope during the early AGB phase}
  \label{profilsabund_endHB}
\end{figure}

\begin{table*}
        \scalebox{1.0}{
        \centering                       
        \begin{tabular}{c | c | c | c | c | c | c | c | c  }      
        \hline\hline                      
        He & \multicolumn{2}{c|}{H} & \multicolumn{2}{c|}{He} & \multicolumn{2}{c|}{[C/N]} & 
        \multicolumn{2}{c}{$^{12}$C/$^{13}$C}  \\  
        (Mass fraction) &  Before & After & Before & After & Before & After & Before & After \\
        \hline                                   
        0.248 & 0.7514 & 0.7396 &    0.2481 & 0.2581 &    0 & 0 &    90.0 & 40.4 \\            
        \hline 
        0.4   & 0.5994 & 0.5905 &    0.4001 & 0.4090 &    -1.30 & -1.36 &    5.65 & 5.33  \\                 
        \hline 
        0.8   & 0.1994 & 0.1994 &    0.8001 & 0.8001 &    -1.61 & -1.61 &    3.99 & 3.99   \\         
        \hline          
        \end{tabular}} 
\caption{Surface abundances of the main elements before and after the first dredge-up on the RGB for the selected 0.8 M$_{\odot}$ models; the values for hydrogen and helium are in mass fraction and those for $^{12}$C/$^{13}$C are in number fraction.}       
\label{table_1}
\end{table*}

\begin{table*}
        \scalebox{1.0}{ 
        \centering                       
        \begin{tabular}{ c | c | c | c | c | c | c | c | c | c | c | c | c | c | c | c | c | c | c}      
        \hline\hline                      
        Y$_{\rm ini}$     & log(L$_\text{bump}$/L$_{\odot}$) & $\tau_\text{bump}$ & log(L$_\text{tip}$/L$_{\odot}$) & log(L/L$_{\odot}$) & log(T$_\text{eff}$) & M$_\text{tot}$ & M$_\text{core}$ & log(T$_\text{eff}$) \\
         & & & & ZAHB & ZAHB (K) & ZAHB (M$_{\odot}$) & ZAHB (M$_{\odot}$) & max-HB (K) \\ \hline 
        0.248 & 2.09 & 3.9    & 3.34 & 1.46 & 4.19 & 0.574 & 0.501 & 4.21 \\ \hline  
        0.26  & 2.11 & 3.6    & 3.33 & 1.48 & 4.17 & 0.583 & 0.498 & 4.18 \\ \hline 
        0.27  & 2.13 & 3.4    & 3.33 & 1.51 & 4.16 & 0.590 & 0.496 & 4.17 \\ \hline 
        0.3   & 2.19 & 2.6    & 3.32 & 1.61 & 4.10 & 0.610 & 0.490 & 4.10 \\ \hline 
        0.33  & 2.25 & 1.9    & 3.30 & 1.73 & 4.02 & 0.629 & 0.482 & 4.07 \\ \hline                
        0.37  & 2.34 & 0.99   & 3.28 & 1.85 & 3.90 & 0.652 & 0.472 & 4.09 \\ \hline 
        0.4   & 2.40 & $\sim$ & 3.26 & 1.92 & 3.85 & 0.668 & 0.463 & 4.14 \\ \hline 
        0.45  & -    & -      & 3.21 & 2.03 & 3.77 & 0.693 & 0.454 & 4.23 \\ \hline 
        0.5   & -    & -      & 3.15 & 2.13 & 3.75 & 0.716 & 0.446 & 4.34 \\ \hline 
        0.55  & -    & -      & 3.06 & 2.24 & 3.75 & 0.735 & 0.443 & 4.44 \\ \hline 
        0.6   & -    & -      & 2.95 & 2.35 & 3.76 & 0.745 & 0.442 & 4.51 \\ \hline 
        0.65  & -    & -      & 2.84 & 2.50 & 3.88 & 0.749 & 0.464 & 4.57 \\ \hline 
        0.7   & -    & -      & 2.66 & 2.65 & 4.04 & 0.755 & 0.490 & 4.64 \\ \hline 
        0.73  & -    & -      & 2.60 & 2.74 & 4.19 & 0.758 & 0.521 & 4.71 \\ \hline 
        0.77  & -    & -      & 2.59 & 2.83 & 4.41 & 0.765 & 0.578 & 4.69 \\ \hline 
        0.8   & -    & -      & 2.66 & 2.76 & 4.57 & 0.773 & 0.627 & 4.74 \\ \hline
        \hline
        \end{tabular}} 
\caption{Luminosity and duration (in Myr) of the RGB bump (as defined in  Fig. \ref{bump_parameters}), and luminosity at the tip of the RGB. Luminosity, effective temperature, total mass, and mass of the core at the ZAHB, and finally maximum effective temperature on the HB for the 0.8 M$_{\odot}$ model as a function of Y$_{\rm ini}$.}         
\label{tablebumpLrgbtip}
\end{table*}
 
\subsection{Horizontal branch}\label{HB}
The  effective temperature and the luminosity of a star on the horizontal branch (HB) directly depend on the total stellar mass, and more importantly on the respective masses of the helium core and of the envelope above the core at central helium ignition, as well as on the opacity in the stellar envelope \citep{Salaris05,Maeder09,Kippen13}. 
For the reasons described above, these quantities strongly vary with the initial helium content for a given initial stellar mass, as shown in Fig.~\ref{Mcoeurenvel} for the 0.8~M$_{\odot}$ case, with consequences on the HB tracks (see Table~\ref{tablebumpLrgbtip} where we give the effective temperature and corresponding luminosity of the 0.8~M$_{\odot}$ models on the ZAHB and at maximum Teff on the HB).
In the same way as for the effective temperature on the ZAHB, one can distinguish two regimes. 
On the one hand, the models that undergo the helium flash (Y$_{\rm ini}$ below $\sim$ 0.5/0.55) arrive on the ZAHB with a higher total mass and a much  lower M$_{core}$/M$_{env}$ ratio when their initial helium content was higher, therefore with a lower T$_{eff}$ and a higher luminosity. 
On the other hand, for the models that do not undergo the helium flash, T$_{eff}$(ZAHB) increases with Y$_{\rm ini }$ because of a higher total mass and an increasing core mass at that phase.
As a result of these two effects, the maximum effective temperature reached on the HB by the 0.8~M$_{\odot}$ models presents a minimum when Y$_{\rm ini} \sim$ 0.33.

In all cases central helium burning occurs in convective conditions (right panels in Fig.~\ref{kippenhahn}).
The resulting CO-core mass increases with increasing Y$_{\rm ini}$ (Fig.~\ref{COcore}; see also Fig.~\ref{profilsabund_endHB} for the abundance profiles inside the three selected 0.8~M$_{\odot}$ models at the end of central helium burning). 

Finally, the lifetime on the horizontal branch is not very sensitive to the initial helium content in the domain where the stars undergo the helium flash (Y$_{\rm ini}$ below $\sim$ 0.5/0.55), since these objects have nearly identical helium-core mass and central temperature at the arrival on the horizontal branch. However, the  lifetime on the horizontal branch drops for the super helium-rich stars. 
For the 0.8~M$_{\odot}$ models with Y$_{\rm ini}$=0.248, 0.4, and 0.8, this phase lasts respectively 94, 91, and 33~Myr (i.e., respectively 0.7\%, 1.7\%, and 12.5\% of the total stellar lifetime).

\subsection{Asymptotic giant branch}\label{AGB}
After central helium exhaustion, all the 0.8~M$_{\odot}$ models with Y$_{\rm ini}$ below or equal to 0.6  reach the TP-AGB; 0.8~M$_{\odot}$ models with higher Y$_{\rm ini}$ spend a short amount of time on the early-AGB before totally losing their envelope. 

During the early-AGB the convective envelope deepens again in mass (except in the extreme Y$_{\rm ini}$=0.8 case that has expelled all its envelope at that time; see Fig.~\ref{envDUP}). No formal second dredge-up occurs, as the convective envelope does not reach the regions where helium abundance increases (see the arrows in Fig.~\ref{profilsabund_endHB}), although it barely reaches the region where the CN-cycle has operated, and the surface abundances of the carbon and nitrogen isotopes slightly change once again. However, this effect is strongly attenuated when Y$_{\rm ini}$ increases, as in the case of the first dredge-up described above.

\begin{figure}[ht]
   \centering
   \includegraphics[width=.47\textwidth]{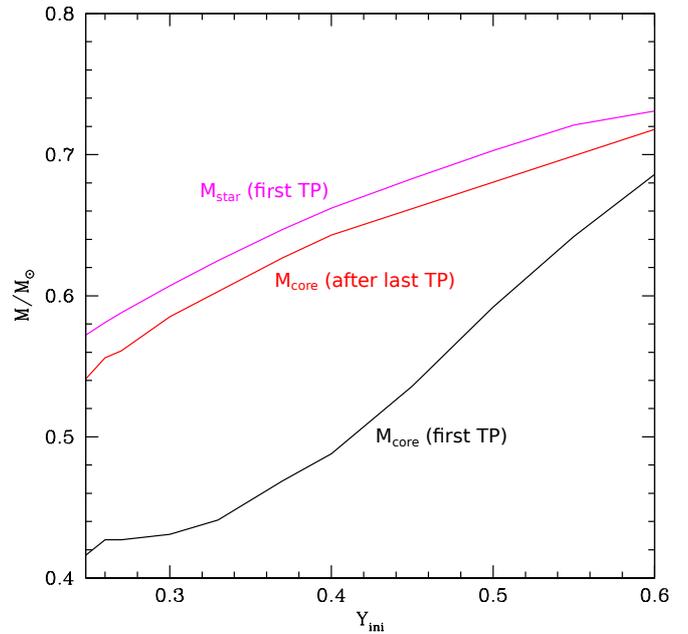}
    \caption{CO core and total stellar mass at the first thermal pulse, and CO core mass after the last one for the 0.8 M$_{\odot}$ models with Y$_{\rm ini}$ between 0.248 and 0.6}
  \label{COcore}
\end{figure}

For the 0.8~M$_{\odot}$ case, the total number of thermal pulses (TPs) first increases with the value of the initial helium content owing to higher stellar mass at the arrival on the TP-AGB. 
However for Y$_{\rm ini}$ above 0.45, this is counterbalanced by the greater ratio between the mass of the CO core and that of the very shallow envelope (Fig.~\ref{COcore}), and the number of thermal pulses then decreases. 
Quantitatively, the 0.8~M$_{\odot}$ with Y$_{\rm ini}$=0.248, 0.4, and 0.6 undergo respectively 8, 18, and 12 thermal pulses, and the TP-AGB phase lasts 1.9, 1.35, and 0.07~Myr, respectively. 

Importantly, for Y$_{\rm ini}$ higher than or equal to 0.65 no thermal pulse occurs and the models spend a very short time on the AGB (post-early AGB) and then evolve directly towards the white dwarf (WD) stage, which strongly reduces the probability of observing super He-rich stars on the AGB phase.

\subsection{White dwarf}\label{WD}

The mass of the degenerate CO-core at  the beginning and the end of the TP-AGB phase is shown in Fig.~\ref{COcore} for the 0.8~M$_{\odot}$ case. 
The mass of the CO white dwarf strongly increases with increasing initial helium content. 
The $Y = 0.248$ model reaches the bluest point of the evolution at 14.3 Gyr, and then starts the final cooling phase (WD cooling curve) with M$_\text{core}  =$  0.570 M$_{\odot}$ whereas for the $Y = 0.4$ model this happens at 5.41 Gyr with M$_\text{core}$ = 0.659 M$_{\odot}$ (Fig.~\ref{COcore}) and at 264 Myr with M$_\text{core}$ = 0.746~M$_{\odot}$ for the $Y = 0.8$ model.
Evolution was not computed further except for a couple of models, since discussion of the WD cooling sequence is beyond the scope of  this paper.


\section{Impact of extreme helium enrichment:  Dependence on the initial stellar mass}
\label{mass_effect}

Here we present the combined effects of varying both initial helium and stellar mass over the considered mass domain between 0.3 and 1.0~$M_{\odot}$. 
We highlight the main  points for specific evolutionary phases, and give the corresponding ranges in (M;Y$_{\rm ini}$) at 10 and 13.4~Gyr that delimit the typical ages of Galactic GCs\footnote{The present models do not include atomic diffusion. Therefore, the ages of 10 and 13.4~Gyr should not be directly compared to those derived in the literature for specific GC from models that include the related processes. Rather, the numbers we give should be compared in terms of relative ages.}. The numbers predicted for (M;Y$_{\rm ini}$) at a given age are expected to vary with [Fe/H] and are given here for the present grid computed with [Fe/H]=-1.75. The various (M;Y$_{\rm ini}$; age) domains 
would also be slightly modified for different prescriptions of the mass-loss rates (we note that in the present computations mass loss is treated according to \citealt{Reimers75} with  $\eta$ = 0.5 on the RGB and to  \citealt{Vassiliadis93} on the AGB). 

\begin{figure}[ht]
   \centering
   \includegraphics[width=.4\textwidth]{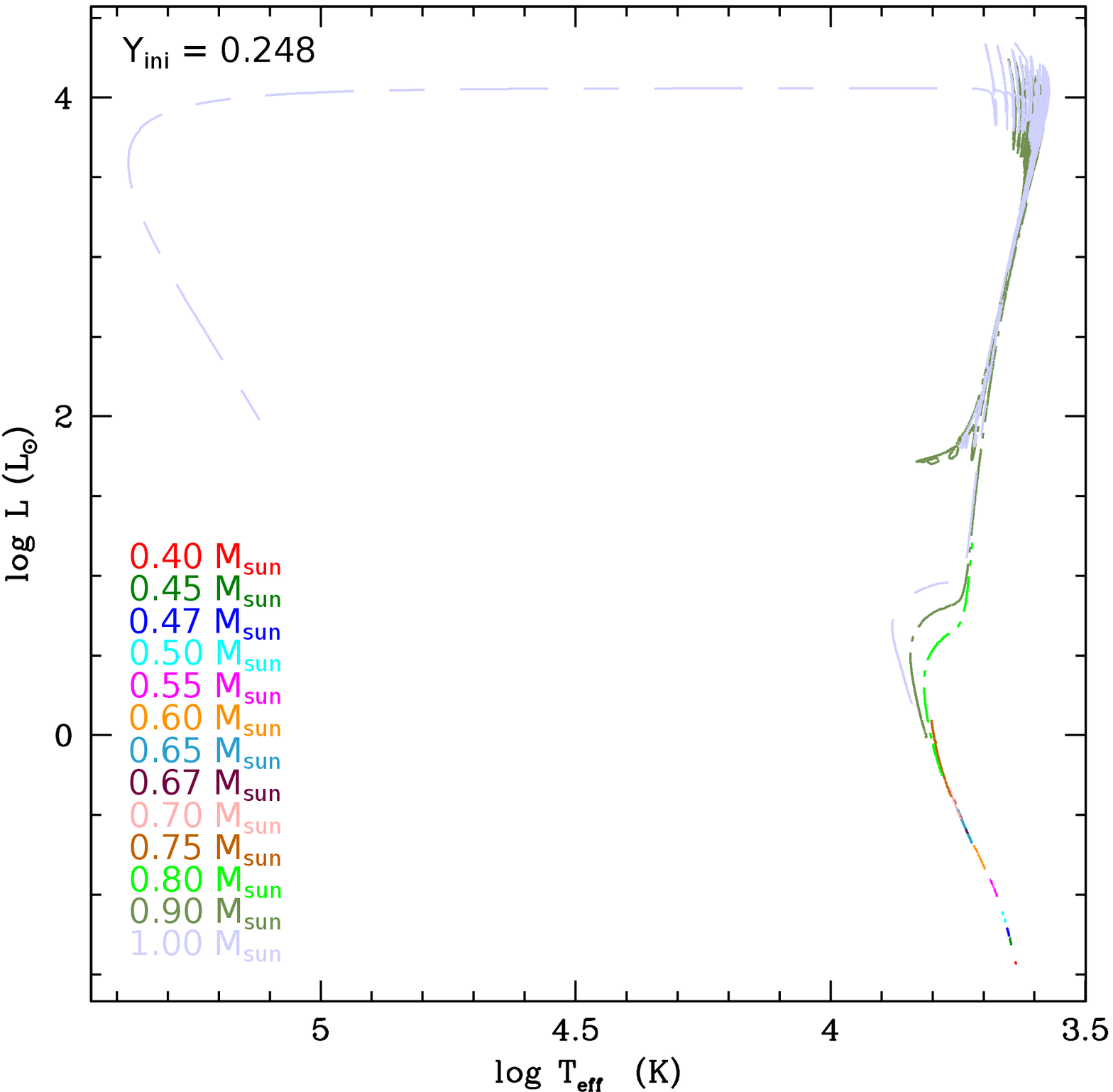}
   \includegraphics[width=.4\textwidth]{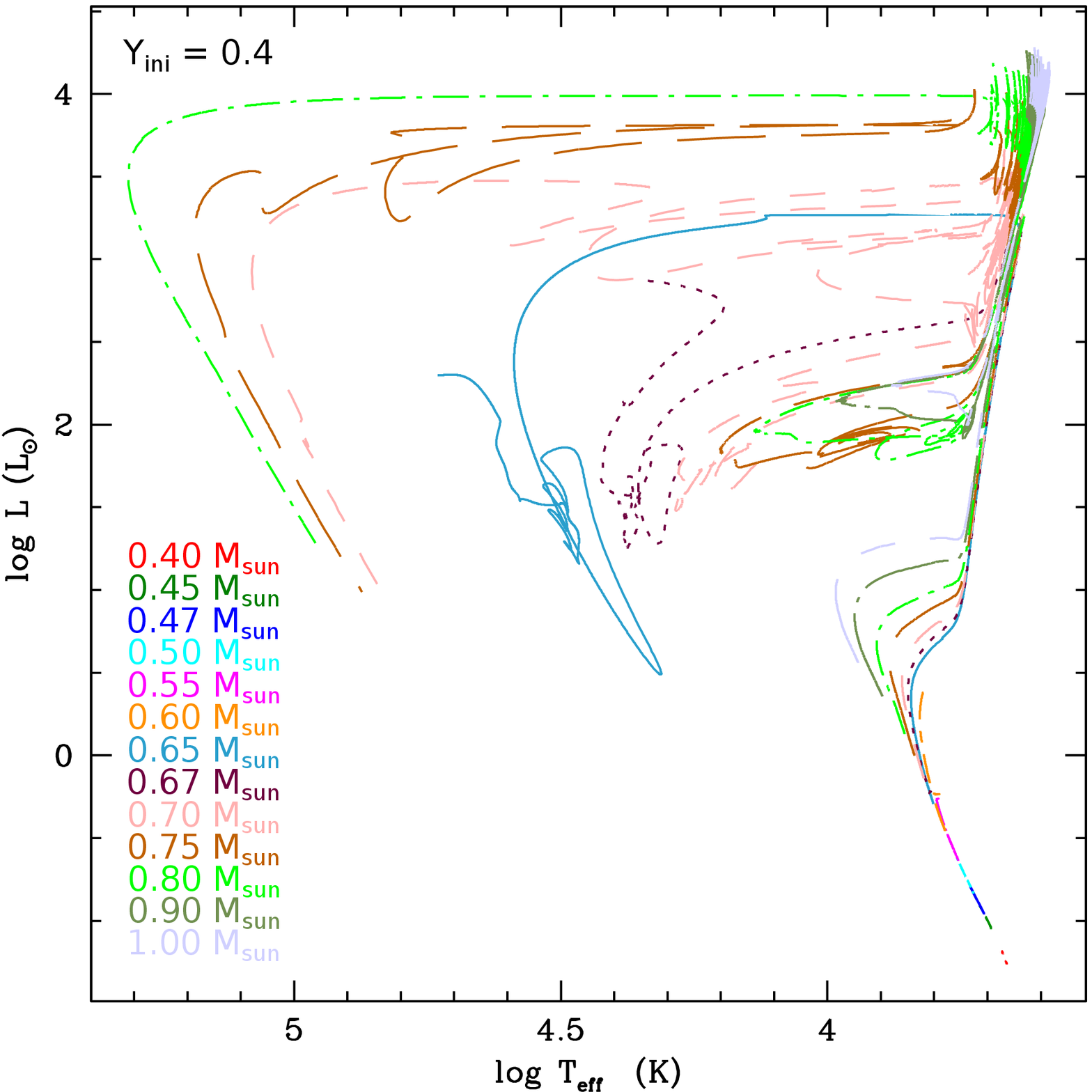}
   \includegraphics[width=.4\textwidth]{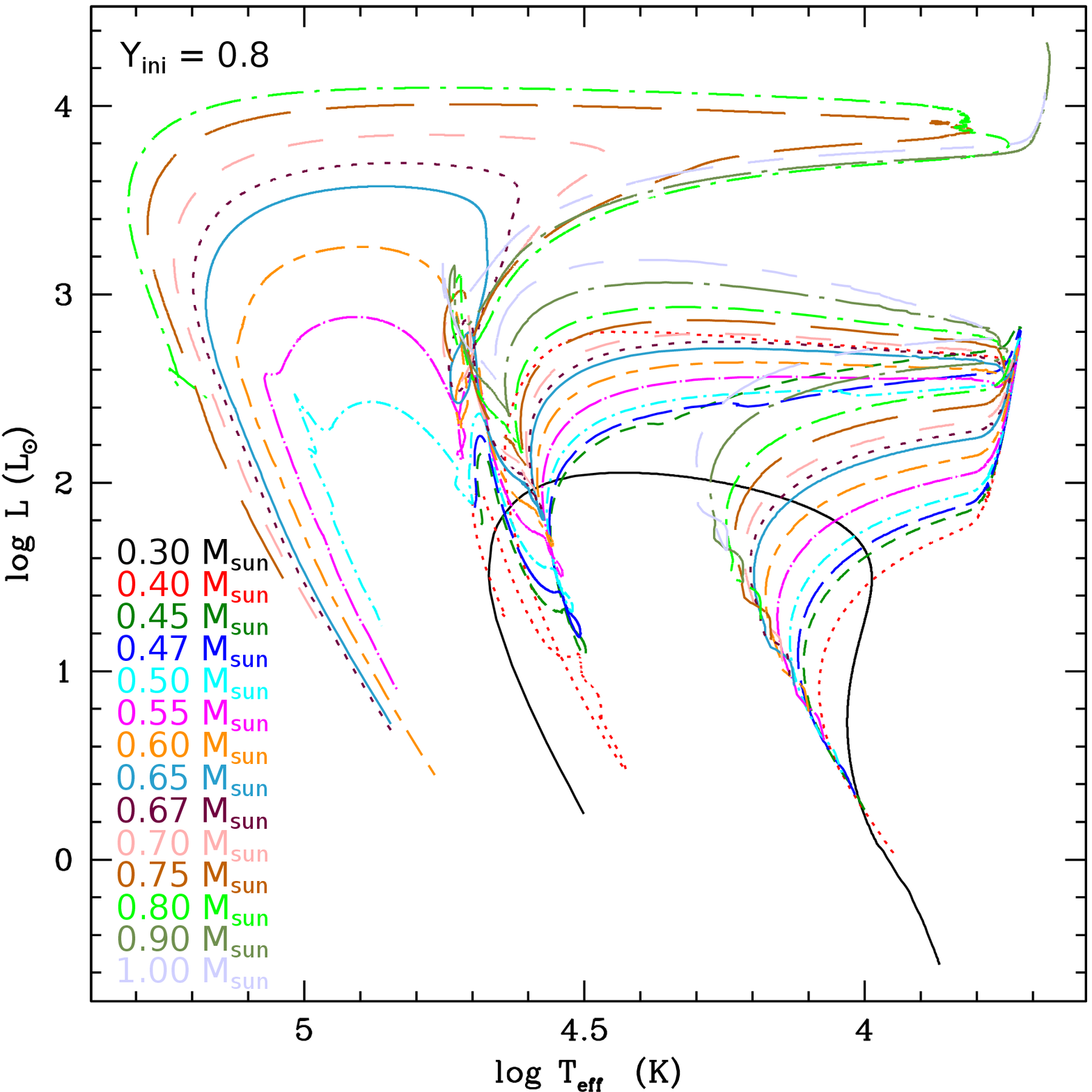}   
    \caption{Evolution in the HRD of all the models with initial stellar masses between 0.3 and 1.0~M$_{\odot}$ for Y$_{\rm ini}$ = 0.248, 0.4, and 0.8 (top , middle, and bottom, respectively). For the very low-mass models, we stop the evolution if it exceeds 14 Gyr. At this age, the (0.3, 0.248) and (0.3; 0.4) models are still on the pre-main sequence, therefore they are not present in the HRD.}
  \label{HRD_Y0p4_0p8}
\end{figure}

\begin{figure*}
   \centering
   \includegraphics[width=.95\textwidth]{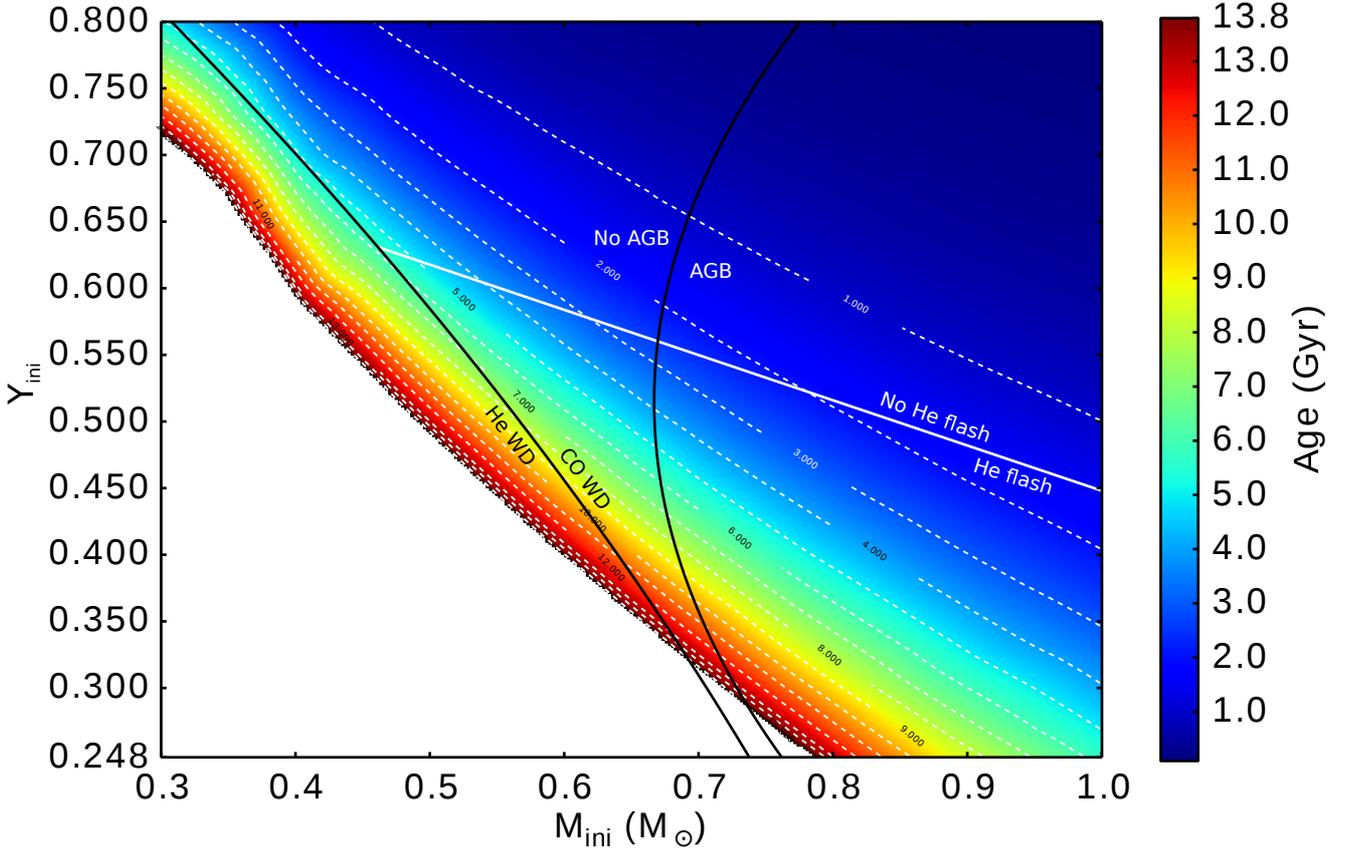}
    \caption{Nature of the white dwarfs and failed-AGB phenomenon as a function of initial stellar mass and Y$_{\rm ini}$. 
    The  black thin line on the left delimits the (M;Y) domains where the stars become either He or CO white dwarfs; it crosses the white line that delimits the domains where helium ignition (when it occurs) starts with a flash or in non-degenerate conditions.
    The curved black thin line separates the stars that climb the AGB from those that do not (failed-AGB). 
    Colors depict the age at the main sequence turn-off, with white dashed lines being isochrones with ages indicated in Gyr; the white area corresponds to the domain where stellar models predict main sequence lifetimes higher than 13.8~Gyr.}
  \label{WDdomain}
\end{figure*}

\noindent {\bf Tracks and lifetime --}
Similarly to  the 0.8~M$_{\odot}$ case, luminosity and effective temperature shifts occur in the Hertzsprung-Russel diagram (HRD) for each given initial mass when Y$_{\rm ini}$ increases (Fig.~\ref{HRD_Y0p4_0p8}), and the duration of each evolutionary phase is reduced accordingly. 
In particular, the main sequence lifetime  strongly depends on Y$_{\rm ini}$ (see the color scale in Fig.~\ref{WDdomain}). 
For a canonical GC age of 13.4~Gyr (see the corresponding isochrone in the figure), the stars that are expected to be at the main sequence turn-off have (M;Y$_{\rm ini}$) between (0.794;0.248) and (0.3;0.719); at 10~Gyr, these quantities are between  (0.858;0.248) and (0.3;0.754).

\begin{figure*}
   \centering
   \includegraphics[width=.95\textwidth]{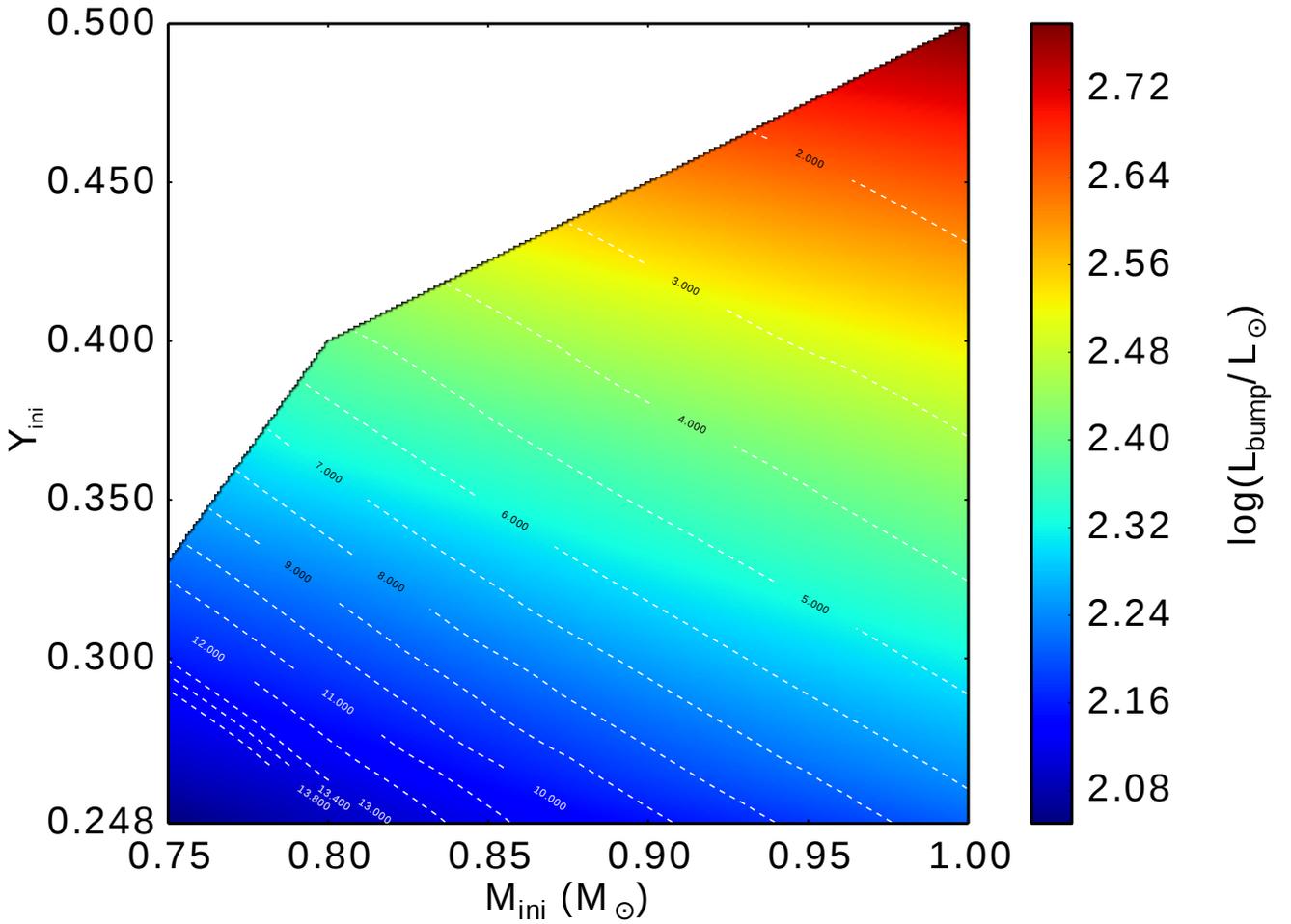}
    \caption{Luminosity of the RGB bump as a function of initial stellar mass and Y$_{\rm ini}$ for those stars that  undergo the bump before 14~Gyr (see ages given in Gyr along the dotted-line isochrones). The white area corresponds to the (M;Y) domain where stars do not undergo the RGB bump (This is also the case  for models with an initial mass lower than 0.75 M$_{\odot}$)}
  \label{bump}
\end{figure*}

\noindent {\bf RGB bump --}
The luminosity of the RGB bump increases with both initial stellar mass and initial helium content (Fig.~\ref{bump}). However, above a certain (M;Y) threshold the stars evolve along the RGB without any luminosity drop, and the bump consequently disappears (white area in Fig.~\ref{bump}) as discussed in \S~\ref{SGB} for the 0.8~M$_{\odot}$ case. 
The stars that are expected to be at the RGB bump have (M;Y$_{\rm ini}$) between (0.813; 0.248) and (0.75; 0.295) 
when we consider an age of 13.4~Gyr (see isochrone), and between (0.881; 0.248) and (0755; 0.336) at 10~Gyr.

\begin{figure*}
\centering
 \includegraphics[width=.95\textwidth]{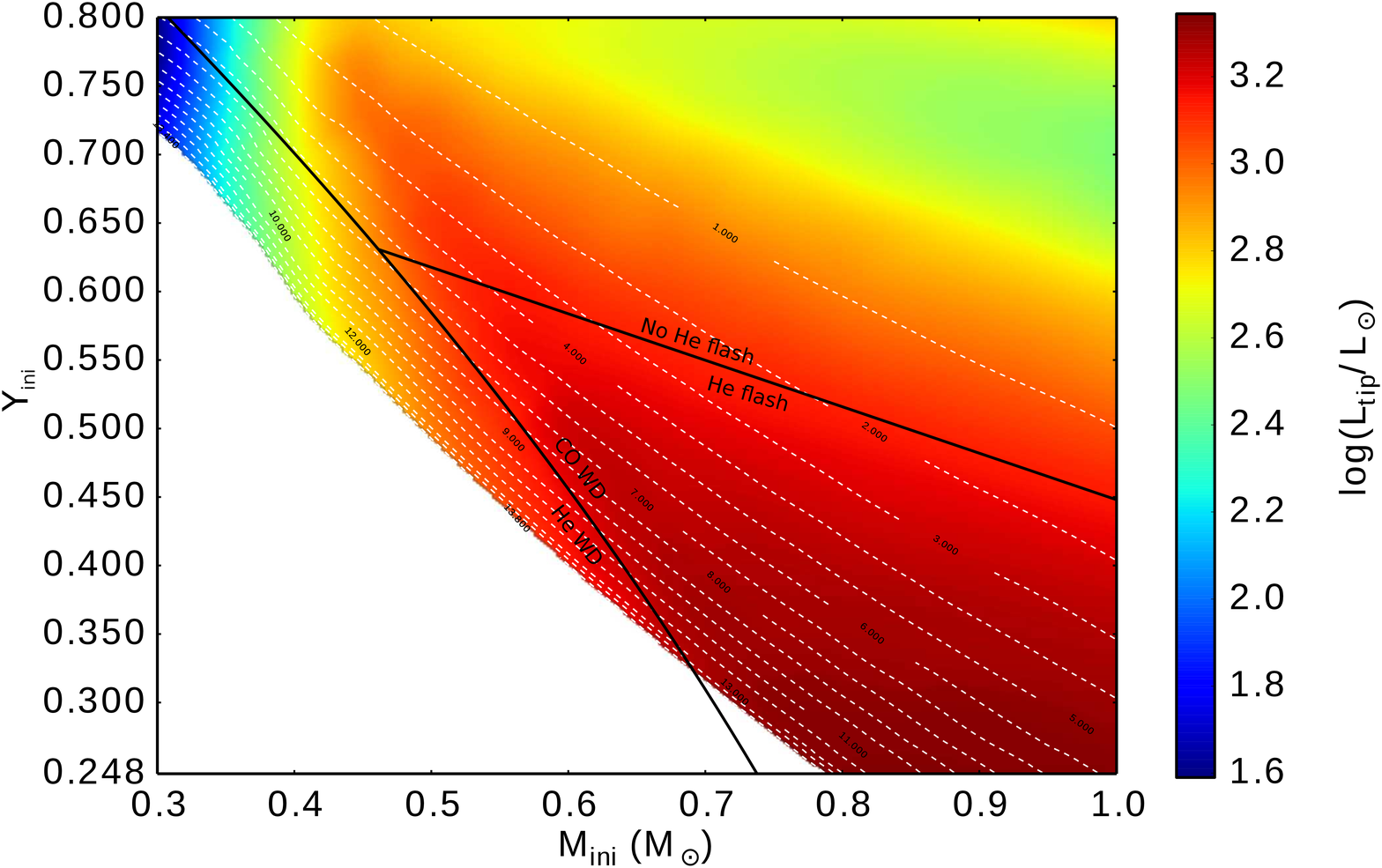}
 \caption{Luminosity of the RGB tip (i.e., when helium ignition occurs) as a function of initial stellar mass and Y$_{\rm ini}$. The black line delimits the domains where helium ignition (when it occurs) starts with a flash or in non-degenerate conditions.
The colors indicate the luminosity of the RGB tip, with white dashed lines being isochrones with MS turn-off ages indicated in Gyr; the white area corresponds to the domain where stellar models predict main sequence lifetimes higher than 13.8~Gyr.}
   
\label{RGBtip_alldomain}
\end{figure*}

\noindent {\bf RGB tip and central helium-burning ignition --} 
The (M;Y$_{\rm ini}$) domain where stars  ignite central helium is indicated in Fig.~\ref{WDdomain} (i.e., in the area found to the right of the curve CO WD / He WD curve).
For the canonical Y$_{\rm ini}$ value of 0.248, all the models of the considered mass range that ignite central helium-burning (i.e., M$_{\rm ini} \geq$ 0.737~M$_{\odot}$) do so in degenerate conditions via the so-called helium flash at the tip of the RGB. 
However, when the initial helium increases, low-mass stars evolve faster on the RGB; they ignite central helium burning smoothly in less degenerate conditions and at a lower luminosity.
The limit in terms of (M;Y$_{\rm ini}$) between the two regimes goes
from (0.46; 0.63) to (1.0; 0.45) for the mass range we are interested in (Fig.~\ref{RGBtip_alldomain}). 

\noindent {\bf Position on the horizontal branch --}
For  a given initial stellar mass, the position on the HB strongly depends on Y$_{\rm ini}$ because it affects whether or not the star ignites He in degenerate conditions, and also depends on the respective masses of the helium core and of the stellar envelope when central helium-burning starts (Sect.~\ref{HB}). In particular, for the models that undergo the He flash, the higher Y$_{\rm ini}$, the hotter the position on the HB. At 13.4~Gyr (and at 10~Gyr, respectively), the models that are in the central helium-burning phases have (M;Y$_{\rm ini}$) between (0.815;0.248) and (0.670;0.355) (respectively between (0.884;0.248) and (0.599;0.458)), 
and the HB covers a domain in T$_{\rm eff}$ between 9550 and 36310 K (respectively between 5370 and 36310/37150 K).

\noindent {\bf TP-AGB vs failed-AGB --}
Among the stars that manage to burn central helium, a significant percentage finish their life without climbing the AGB. 
The corresponding (M;Y$_{\rm ini}$) domain of these so-called failed-AGB (see, e.g., \citealt{Greggio90}) is shown in Fig.~\ref{WDdomain} (left of the No AGB / AGB line).
These stars undergo central helium-burning at very high effective temperatures on the HB; 
there the stellar envelope surrounding the nuclear active core is extremely thin. Consequently, once helium fuel is exhausted at the center, the energy released by the contraction of the core and by hydrogen shell-burning mainly diffuses across the small, transparent helium-rich external layers and fails to swell them up. Therefore, the stars  first evolve at constant effective temperature towards higher luminosity before moving directly towards the white dwarf region without passing through the AGB.
This behavior has been proposed as a suitable explanation for the lack of sodium-rich stars on the AGB in NGC 6752 \citep{Charbonnel13}.
At 13.4~Gyr (and at 10~Gyr respectively), the models that are on the AGB have (M;Y$_{\rm ini}$) between (0.815;0.248) and (0.67;0.355) (respectively between (0.884;0.248) and (0.599;0.458)).

\noindent {\bf Nature of the white dwarfs --} The composition (He vs CO) of the stellar remnant strongly depends on the initial (M;Y) as is shown in Fig.~\ref{WDdomain}.
In the case of canonical helium abundance no He-WDs are expected to be present in GCs today because the stellar lifetimes of the corresponding low-mass progenitors are much longer than the Hubble time\footnote{For Y$_{\rm ini}$ = 0.248 and $Z = 5 \times 10^{-4}$, the maximum initial mass of a He-WD progenitor is 0.738 M$_{\odot}$, which has a turn-off age of $\sim$ 17.6 Gyr. Therefore, for this composition He-WD might be expected only from exotic evolution paths; this can happen for example if the low-mass progenitors have undergone extreme mass-loss events, e.g., in close binary systems \citep[e.g.,][]{Webbink75,Strickler09}.}.
However He-WDs are expected to form on much shorter lifetimes for increased initial helium abundance above $\sim$0.33; in addition, the higher Y$_{\rm ini}$, the lower the initial mass for the star to become a He-WD. 
The consequences of this behavior on the chemical properties of GC host WDs will be investigated in detail in a forthcoming paper.

\section{Conclusions}\label{conclusion}

This paper presents the first grid of models of low-mass stars (between 0.3 and 1~M$_{\odot}$) with initial chemical composition derived from the ejecta of the fast rotating massive star models of \cite{Decressin07a,Decressin07b} at [Fe/H]=-1.75 mixed with original proto-GC gas. 
The initial helium mass fraction varies between 0.248 and 0.8, and the initial abundances of C, N, O, Na, Mg, and Al change accordingly. 

The main properties of the models are presented as a function of initial mass and initial helium abundance. 
We discuss the effects of a very high initial helium content 
on the evolution paths in the HRD, on the duration and characteristics of the main evolutionary phases, and on the chemical nature of the WD remnants. 
We also give the predicted ranges in initial mass and helium content of the stars in different locations of the HRD at ages of 10 and 13.4~Gyr that delimit the age range of Galactic GCs.

This grid of models has been successfully used to explain the lack of Na-rich AGB stars in the GC NGC~6752 \citep{Charbonnel14}. Other aspects related to the properties of stellar populations in GCs and of the GC CMDs at various phases of their evolution will be investigated in forthcoming papers in this series.


\section*{Appendix -- Content of electronic tables for the grid}

We provide files containing relevant evolution characteristics (Table~\ref{tab:gridtable}) for the 224 models of the present grid. 
All data detailed in Table \ref{tab:gridtable} are available on the website \url{http://obswww.unige.ch/Recherche/evol/-Database}.
For each mass and initial helium content, models predictions are provided from the beginning of the pre-main sequence (along the Hayashi track) up to the brightest point of the evolution on the AGB. 
For each model, we have selected 500 points to allow a good description of the full raw tracks. 
First, the most important evolution keypoints are determined: 
\begin{enumerate}
\item the beginning of the pre-main sequence;
\item the zero age main-sequence defined as the time when the central
  hydrogen abundance has decreased by 0.003 in mass fraction compared to its initial value.
\item the turning point when the central mass fraction of hydrogen is equal to 0.06 on the main sequence;
\item the end of core H-burning 
\item the bottom of the red giant branch (RGB);
\item the RGB-tip;
\item the local minimum of luminosity during central He-burning;
\item the local maximum of $T_\text{eff}$ during central He-burning;
\item the bottom of AGB: the point with the local minimum of luminosity after the loop on HB 
\item the brightest point on the AGB.
\end{enumerate}

Then, we chose points from the results of evolutionary computation as follows: 
\begin{itemize}
\item 100 points are determined alongside the pre-main sequence with constant time-steps.
The value of each quantity given in Table \ref{tab:gridtable} is computed from their values taken from the two nearest models calculated with STAREVOL.
\item The same procedure is applied between points 2 and 3 and  points 3 and 4 where we use step in X$_{C}$ instead of time; 85 and 25 points are selected, respectively.
\item Between keypoints 4 and 5 we interpolate 60 points with equal time steps. 
\item Then 20 points are determined between keypoints 5 and 6 with equal intervals of $\log L$.
\item 20, 70, and 70 points are selected with equal central hydrogen abundance $X_{C}$ steps between
keypoints 6 and 7, keypoints 7 and 8,  and keypoints 8 and 9. 
\item 50 points are selected between keypoints 9 and
10 with the same $\log L$ step on the AGB. 
\end{itemize}

\begin{table*}[htbf]
\hspace{3cm}
  \caption{Description of table containing the results of our evolution models.}
   \centering 
  \scalebox{0.92}{
  \begin{tabular}{|| l || l || l ||}
  \hline \hline
  \bf{Stellar parameters}                                            & \bf{Surface abundances}                       & \bf{Central abundances}                          \\
  \hline \hline
 - Model number                                                                       &  $^{1}H$ $^{2}H$      &   $^{1}H$          \\
 - Maximum  temperature $T_{max}$      (K)   &              $^{3}He$ $^{4}He$                                               & $^{3}He$ $^{4}He$                \\
 - Mass coordinate of $T_{max}$    (\Ms)     &            $^{6}Li$ $^{7}Li$                                                    &                      \\
 - Effective temperature T$_{eff}$ (K)                      &     $^{7}Be$ $^{9}Be$                        &                              \\
 - Surface luminosity L (L$_{\odot}$)                                       &     $^{10}B$ $^{11}B$                     &                  \\
 - Photospheric radius radius R$_{eff}$ (R$_{\odot}$)                   &$^{12}C$ $^{13}C$ $^{14}C$                             &   $^{12}C$ $^{13}C$ $^{14}C$                   \\
 - Photospheric density $\rho_{eff}$ (g.$cm^{-3}$)       &        $^{14}N$ $^{15}N$                     &            $^{14}N$           \\
 - Density at the location of $T_{max}$, $\rho_{max}$ (g.$cm^{-3}$)        &$^{16}O$  $^{17}O$ $^{18}O$                          &       $^{16}O$  $^{17}O$ $^{18}O$                \\
 - Stellar mass M (\Ms)                                  &  $^{19}F$                              &         $^{19}F$       \\
 - Mass-loss rate ($\Ms.yr^{-1}$)                          &$^{20}Ne$ $^{21}Ne$ $^{22}Ne$   &     $^{20}Ne$ $^{21}Ne$ $^{22}Ne$                 \\
 - Age t (yr)                                                                          &  $^{23}Na$     &      $^{23}Na$      \\
 - Photospheric gravity $log (g_{eff})$  (log(cgs))  &  $^{24}Mg$ $^{25}Mg$ $^{26}Mg$                         &     $^{24}Mg$ $^{25}Mg$ $^{26}Mg$             \\
 - Central temperature $T_{c}$ (K)    &  $^{26}Al$ $^{27}Al$       &     $^{26}Al$ $^{27}Al$     \\
 - Central pressure $P_{c}$     &  $^{28}Si$                                            &     $^{28}Si$              \\
 - Surface velocity $v_{surf}$ (km.s$^{-1}$) &&\\ 
 &&\\
 - Mass at the base of convective envelope (\Ms)        &&\\
 - Mass of the core (\Ms)        &&\\
  \hline \hline  
  \end{tabular}}
  \label{tab:gridtable}
\end{table*}

\begin{acknowledgements}
We thank G. Meynet for fruitful discussions and the anonymous referee for constructive suggestions on the manuscript. We acknowledge financial support from the Swiss National Science Foundation (FNS) for the project 200020-140346 "New perspectives on the chemical and dynamical evolution of globular clusters in light of new generation stellar models" (PI C.C.). We thank the International Space Science Institute (ISSI, Bern, CH) for welcoming the activities of ISSI Team 271 ``Massive Star Clusters across the Hubble Time" (2013 - 2015; team leader C.C.).
T.D. acknowledges financial support from the UE Programme (FP7/2007-2013) under grant  No. 267251 “Astronomy Fellowships in Italy” (ASTROFit).
\end{acknowledgements}
\bibliographystyle{aa}
\bibliography{Bibliography}

\begin{thebibliography}{84}
\expandafter\ifx\csname natexlab\endcsname\relax\def\natexlab#1{#1}\fi

\bibitem[{{Bastian} {et~al.}(2013{\natexlab{a}}){Bastian}, {Cabrera-Ziri},
  {Davies}, \& {Larsen}}]{Bastianetal2013b}
{Bastian}, N., {Cabrera-Ziri}, I., {Davies}, B., \& {Larsen}, S.~S.
  2013{\natexlab{a}}, \mnras, 436, 2852

\bibitem[{{Bastian} {et~al.}(2014){Bastian}, {Hollyhead}, \&
  {Cabrera-Ziri}}]{Bastianetal14}
{Bastian}, N., {Hollyhead}, K., \& {Cabrera-Ziri}, I. 2014, \mnras, 445, 378

\bibitem[{{Bastian} {et~al.}(2013{\natexlab{b}}){Bastian}, {Lamers}, {de Mink},
  {Longmore}, {Goodwin}, \& {Gieles}}]{Bastianetal2013a}
{Bastian}, N., {Lamers}, H.~J.~G.~L.~M., {de Mink}, S.~E., {et~al.}
  2013{\natexlab{b}}, \mnras, 436, 2398

\bibitem[{{Bastian} \& {Strader}(2014)}]{BastianStrader14}
{Bastian}, N. \& {Strader}, J. 2014, \mnras, 443, 3594

\bibitem[{{Bedin} {et~al.}(2004){Bedin}, {Piotto}, {Anderson}, {Cassisi},
  {King}, {Momany}, \& {Carraro}}]{Bedin04}
{Bedin}, L.~R., {Piotto}, G., {Anderson}, J., {et~al.} 2004, \apjl, 605, L125

\bibitem[{{Bragaglia} {et~al.}(2014){Bragaglia}, {Sneden}, {Carretta},
  {Gratton}, {Lucatello}, {Bernath}, {Brooke}, \& {Ram}}]{Bragagliaetal14}
{Bragaglia}, A., {Sneden}, C., {Carretta}, E., {et~al.} 2014, \apj, 796, 68

\bibitem[{{Cabrera-Ziri} {et~al.}(2014){Cabrera-Ziri}, {Bastian}, {Davies},
  {Magris}, {Bruzual}, \& {Schweizer}}]{CabreraZirietal2014}
{Cabrera-Ziri}, I., {Bastian}, N., {Davies}, B., {et~al.} 2014, \mnras, 441,
  2754

\bibitem[{{Caloi} \& {D'Antona}(2007)}]{Caloi07}
{Caloi}, V. \& {D'Antona}, F. 2007, \aap, 463, 949

\bibitem[{{Campbell} {et~al.}(2013){Campbell}, {D'Orazi}, {Yong},
  {Constantino}, {Lattanzio}, {Stancliffe}, {Angelou}, {Wylie-de Boer}, \&
  {Grundahl}}]{Campbell13}
{Campbell}, S.~W., {D'Orazi}, V., {Yong}, D., {et~al.} 2013, \nat, 498, 198

\bibitem[{{Carretta}(2013)}]{Carretta13}
{Carretta}, E. 2013, \aap, 557, A128

\bibitem[{{Carretta} {et~al.}(2009{\natexlab{a}}){Carretta}, {Bragaglia},
  {Gratton}, {D'Orazi}, \& {Lucatello}}]{Carretta09c}
{Carretta}, E., {Bragaglia}, A., {Gratton}, R., {D'Orazi}, V., \& {Lucatello},
  S. 2009{\natexlab{a}}, \aap, 508, 695

\bibitem[{{Carretta} {et~al.}(2009{\natexlab{b}}){Carretta}, {Bragaglia},
  {Gratton}, {Lucatello}, {Catanzaro}, {Leone}, {Bellazzini}, {Claudi},
  {D'Orazi}, {Momany}, {Ortolani}, {Pancino}, {Piotto}, {Recio-Blanco}, \&
  {Sabbi}}]{Carretta09b}
{Carretta}, E., {Bragaglia}, A., {Gratton}, R.~G., {et~al.} 2009{\natexlab{b}},
  \aap, 505, 117

\bibitem[{{Carretta} {et~al.}(2012){Carretta}, {Bragaglia}, {Gratton},
  {Lucatello}, \& {D'Orazi}}]{Carretta12}
{Carretta}, E., {Bragaglia}, A., {Gratton}, R.~G., {Lucatello}, S., \&
  {D'Orazi}, V. 2012, \apjl, 750, L14

\bibitem[{{Carretta} {et~al.}(2007){Carretta}, {Bragaglia}, {Gratton},
  {Lucatello}, \& {Momany}}]{Carretta07a}
{Carretta}, E., {Bragaglia}, A., {Gratton}, R.~G., {Lucatello}, S., \&
  {Momany}, Y. 2007, \aap, 464, 927

\bibitem[{{Carretta} {et~al.}(2010){Carretta}, {Bragaglia}, {Gratton},
  {Recio-Blanco}, {Lucatello}, {D'Orazi}, \& {Cassisi}}]{Carretta10}
{Carretta}, E., {Bragaglia}, A., {Gratton}, R.~G., {et~al.} 2010, \aap, 516,
  A55

\bibitem[{{Carretta} {et~al.}(2005){Carretta}, {Gratton}, {Lucatello},
  {Bragaglia}, \& {Bonifacio}}]{Carretta05}
{Carretta}, E., {Gratton}, R.~G., {Lucatello}, S., {Bragaglia}, A., \&
  {Bonifacio}, P. 2005, \aap, 433, 597

\bibitem[{{Cassisi} {et~al.}(2013){Cassisi}, {Mucciarelli}, {Pietrinferni},
  {Salaris}, \& {Ferguson}}]{Cassisi13}
{Cassisi}, S., {Mucciarelli}, A., {Pietrinferni}, A., {Salaris}, M., \&
  {Ferguson}, J. 2013, \aap, 554, A19

\bibitem[{{Cassisi} \& {Salaris}(2014)}]{CassisiSalaris2014}
{Cassisi}, S. \& {Salaris}, M. 2014, \aap, 563, A10

\bibitem[{{Charbonnel} {et~al.}(2013){Charbonnel}, {Chantereau}, {Decressin},
  {Meynet}, \& {Schaerer}}]{Charbonnel13}
{Charbonnel}, C., {Chantereau}, W., {Decressin}, T., {Meynet}, G., \&
  {Schaerer}, D. 2013, \aap, 557, L17

\bibitem[{{Charbonnel} {et~al.}(2014){Charbonnel}, {Chantereau}, {Krause},
  {Primas}, \& {Wang}}]{Charbonnel14}
{Charbonnel}, C., {Chantereau}, W., {Krause}, M., {Primas}, F., \& {Wang}, Y.
  2014, \aap, 569, L6

\bibitem[{{Coc} {et~al.}(2013){Coc}, {Uzan}, \& {Vangioni}}]{Coc13}
{Coc}, A., {Uzan}, J.-P., \& {Vangioni}, E. 2013, ArXiv e-prints

\bibitem[{{Dalessandro} {et~al.}(2011){Dalessandro}, {Salaris}, {Ferraro},
  {Cassisi}, {Lanzoni}, {Rood}, {Fusi Pecci}, \& {Sabbi}}]{Dalessandro11}
{Dalessandro}, E., {Salaris}, M., {Ferraro}, F.~R., {et~al.} 2011, \mnras, 410,
  694

\bibitem[{{D'Antona} {et~al.}(2005){D'Antona}, {Bellazzini}, {Caloi}, {Pecci},
  {Galleti}, \& {Rood}}]{D'Antona05}
{D'Antona}, F., {Bellazzini}, M., {Caloi}, V., {et~al.} 2005, \apj, 631, 868

\bibitem[{{D'Antona} \& {Caloi}(2004)}]{D'Antona04}
{D'Antona}, F. \& {Caloi}, V. 2004, \apj, 611, 871

\bibitem[{{D'Antona} {et~al.}(2002){D'Antona}, {Caloi}, {Montalb{\'a}n},
  {Ventura}, \& {Gratton}}]{D'Antona02}
{D'Antona}, F., {Caloi}, V., {Montalb{\'a}n}, J., {Ventura}, P., \& {Gratton},
  R. 2002, \aap, 395, 69

\bibitem[{{D'Antona} {et~al.}(2010){D'Antona}, {Caloi}, \&
  {Ventura}}]{D'Antona10}
{D'Antona}, F., {Caloi}, V., \& {Ventura}, P. 2010, \mnras, 405, 2295

\bibitem[{{D'Antona} {et~al.}(2014){D'Antona}, {Ventura}, {Decressin},
  {Vesperini}, \& {D'Ercole}}]{D'Antona14}
{D'Antona}, F., {Ventura}, P., {Decressin}, T., {Vesperini}, E., \& {D'Ercole},
  A. 2014, \mnras, 443, 3302

\bibitem[{{de Mink} {et~al.}(2009){de Mink}, {Pols}, {Langer}, \&
  {Izzard}}]{DeMink09}
{de Mink}, S.~E., {Pols}, O.~R., {Langer}, N., \& {Izzard}, R.~G. 2009, \aap,
  507, L1

\bibitem[{{Decressin} {et~al.}(2007{\natexlab{a}}){Decressin}, {Charbonnel}, \&
  {Meynet}}]{Decressin07a}
{Decressin}, T., {Charbonnel}, C., \& {Meynet}, G. 2007{\natexlab{a}}, \aap,
  475, 859

\bibitem[{{Decressin} {et~al.}(2007{\natexlab{b}}){Decressin}, {Meynet},
  {Charbonnel}, {Prantzos}, \& {Ekstr{\"o}m}}]{Decressin07b}
{Decressin}, T., {Meynet}, G., {Charbonnel}, C., {Prantzos}, N., \&
  {Ekstr{\"o}m}, S. 2007{\natexlab{b}}, \aap, 464, 1029

\bibitem[{{Demarque} {et~al.}(1971){Demarque}, {Mengel}, \&
  {Aizenman}}]{Demarque71}
{Demarque}, P., {Mengel}, J.~G., \& {Aizenman}, M.~L. 1971, \apj, 163, 37

\bibitem[{{Denissenkov} \& {Hartwick}(2014)}]{DenissenkovHartwick14}
{Denissenkov}, P.~A. \& {Hartwick}, F.~D.~A. 2014, \mnras, 437, L21

\bibitem[{{D'Ercole} {et~al.}(2012){D'Ercole}, {D'Antona}, {Carini},
  {Vesperini}, \& {Ventura}}]{D'ercole12}
{D'Ercole}, A., {D'Antona}, F., {Carini}, R., {Vesperini}, E., \& {Ventura}, P.
  2012, \mnras, 423, 1521

\bibitem[{{D'Ercole} {et~al.}(2010){D'Ercole}, {D'Antona}, {Ventura},
  {Vesperini}, \& {McMillan}}]{D'ercole10}
{D'Ercole}, A., {D'Antona}, F., {Ventura}, P., {Vesperini}, E., \& {McMillan},
  S.~L.~W. 2010, \mnras, 407, 854

\bibitem[{{D'Ercole} {et~al.}(2011){D'Ercole}, {D'Antona}, \&
  {Vesperini}}]{D'ercole11}
{D'Ercole}, A., {D'Antona}, F., \& {Vesperini}, E. 2011, \mnras, 415, 1304

\bibitem[{{Doherty} {et~al.}(2014){Doherty}, {Gil-Pons}, {Lau}, {Lattanzio}, \&
  {Siess}}]{Doherty14}
{Doherty}, C.~L., {Gil-Pons}, P., {Lau}, H.~H.~B., {Lattanzio}, J.~C., \&
  {Siess}, L. 2014, \mnras, 437, 195

\bibitem[{{Dotter} {et~al.}(2010){Dotter}, {Sarajedini}, {Anderson},
  {Aparicio}, {Bedin}, {Chaboyer}, {Majewski}, {Mar{\'{\i}}n-Franch}, {Milone},
  {Paust}, {Piotto}, {Reid}, {Rosenberg}, \& {Siegel}}]{Dotter10}
{Dotter}, A., {Sarajedini}, A., {Anderson}, J., {et~al.} 2010, \apj, 708, 698

\bibitem[{{Forestini} \& {Charbonnel}(1997)}]{Forestini97}
{Forestini}, M. \& {Charbonnel}, C. 1997, \aaps, 123, 241

\bibitem[{{Gratton} {et~al.}(2010){Gratton}, {Carretta}, {Bragaglia},
  {Lucatello}, \& {D'Orazi}}]{Gratton10}
{Gratton}, R.~G., {Carretta}, E., {Bragaglia}, A., {Lucatello}, S., \&
  {D'Orazi}, V. 2010, \aap, 517, A81

\bibitem[{{Greggio} \& {Renzini}(1990)}]{Greggio90}
{Greggio}, L. \& {Renzini}, A. 1990, \apj, 364, 35

\bibitem[{{Grevesse} \& {Noels}(1993)}]{Grevesse93}
{Grevesse}, N. \& {Noels}, A. 1993, in Origin and Evolution of the Elements,
  ed. N.~{Prantzos}, E.~{Vangioni-Flam}, \& M.~{Casse}, 15--25

\bibitem[{{Iben} \& {Rood}(1969)}]{IbenRood1969}
{Iben}, I. \& {Rood}, R.~T. 1969, \nat, 223, 933

\bibitem[{{Iben} \& {Faulkner}(1968)}]{Iben68}
{Iben}, Jr., I. \& {Faulkner}, J. 1968, \apj, 153, 101

\bibitem[{{Izzard} {et~al.}(2013){Izzard}, {de Mink}, {Pols}, {Langer}, {Sana},
  \& {de Koter}}]{Izzard13}
{Izzard}, R.~G., {de Mink}, S.~E., {Pols}, O.~R., {et~al.} 2013, \memsai, 84,
  171

\bibitem[{{King} {et~al.}(2012){King}, {Bedin}, {Cassisi}, {Milone}, {Bellini},
  {Piotto}, {Anderson}, {Pietrinferni}, \& {Cordier}}]{King12}
{King}, I.~R., {Bedin}, L.~R., {Cassisi}, S., {et~al.} 2012, \aj, 144, 5

\bibitem[{{Kippenhahn} {et~al.}(2013){Kippenhahn}, {Weigert}, \&
  {Weiss}}]{Kippen13}
{Kippenhahn}, R., {Weigert}, A., \& {Weiss}, A. 2013, {Stellar Structure and
  Evolution}

\bibitem[{{Krause} {et~al.}(2013){Krause}, {Charbonnel}, {Decressin}, {Meynet},
  \& {Prantzos}}]{Krause13}
{Krause}, M., {Charbonnel}, C., {Decressin}, T., {Meynet}, G., \& {Prantzos},
  N. 2013, \aap, 552, A121

\bibitem[{{Kruijssen}(2014)}]{Kruijssen2014}
{Kruijssen}, J.~M.~D. 2014, Classical and Quantum Gravity, 31, 244006

\bibitem[{{Lagarde} {et~al.}(2012){Lagarde}, {Decressin}, {Charbonnel},
  {Eggenberger}, {Ekstr{\"o}m}, \& {Palacios}}]{Lagarde12}
{Lagarde}, N., {Decressin}, T., {Charbonnel}, C., {et~al.} 2012, \aap, 543,
  A108

\bibitem[{{MacLean} {et~al.}(2015){MacLean}, {De Silva}, \&
  {Lattanzio}}]{Maclean15}
{MacLean}, B.~T., {De Silva}, G.~M., \& {Lattanzio}, J. 2015, \mnras, 446, 3556

\bibitem[{{Maeder}(2009)}]{Maeder09}
{Maeder}, A. 2009, {Physics, Formation and Evolution of Rotating Stars}

\bibitem[{{Maeder} \& {Meynet}(2006)}]{Maeder06}
{Maeder}, A. \& {Meynet}, G. 2006, \aap, 448, L37

\bibitem[{{Marino} {et~al.}(2014){Marino}, {Milone}, {Przybilla}, {Bergemann},
  {Lind}, {Asplund}, {Cassisi}, {Catelan}, {Casagrande}, {Valcarce}, {Bedin},
  {Cort{\'e}s}, {D'Antona}, {Jerjen}, {Piotto}, {Schlesinger}, {Zoccali}, \&
  {Angeloni}}]{Marino14}
{Marino}, A.~F., {Milone}, A.~P., {Przybilla}, N., {et~al.} 2014, \mnras, 437,
  1609

\bibitem[{{Martell} \& {Grebel}(2010)}]{Martell10}
{Martell}, S.~L. \& {Grebel}, E.~K. 2010, \aap, 519, A14

\bibitem[{{Milone} {et~al.}(2008){Milone}, {Bedin}, {Piotto}, {Anderson},
  {King}, {Sarajedini}, {Dotter}, {Chaboyer}, {Mar{\'{\i}}n-Franch},
  {Majewski}, {Aparicio}, {Hempel}, {Paust}, {Reid}, {Rosenberg}, \&
  {Siegel}}]{Milone08}
{Milone}, A.~P., {Bedin}, L.~R., {Piotto}, G., {et~al.} 2008, \apj, 673, 241

\bibitem[{{Milone} {et~al.}(2013){Milone}, {Marino}, {Piotto}, {Bedin},
  {Anderson}, {Aparicio}, {Bellini}, {Cassisi}, {D'Antona}, {Grundahl},
  {Monelli}, \& {Yong}}]{Milone13}
{Milone}, A.~P., {Marino}, A.~F., {Piotto}, G., {et~al.} 2013, \apj, 767, 120

\bibitem[{{Milone} {et~al.}(2010){Milone}, {Piotto}, {King}, {Bedin},
  {Anderson}, {Marino}, {Momany}, {Malavolta}, \& {Villanova}}]{Milone10}
{Milone}, A.~P., {Piotto}, G., {King}, I.~R., {et~al.} 2010, \apj, 709, 1183

\bibitem[{{Mowlavi} {et~al.}(1998){Mowlavi}, {Meynet}, {Maeder}, {Schaerer}, \&
  {Charbonnel}}]{Mowlavi98}
{Mowlavi}, N., {Meynet}, G., {Maeder}, A., {Schaerer}, D., \& {Charbonnel}, C.
  1998, \aap, 335, 573

\bibitem[{{Palacios} {et~al.}(2006){Palacios}, {Charbonnel}, {Talon}, \&
  {Siess}}]{Palacios06}
{Palacios}, A., {Charbonnel}, C., {Talon}, S., \& {Siess}, L. 2006, \aap, 453,
  261

\bibitem[{{Pasquini} {et~al.}(2005){Pasquini}, {Bonifacio}, {Molaro},
  {Francois}, {Spite}, {Gratton}, {Carretta}, \& {Wolff}}]{Pasquini05}
{Pasquini}, L., {Bonifacio}, P., {Molaro}, P., {et~al.} 2005, \aap, 441, 549

\bibitem[{{Pietrinferni} {et~al.}(2009){Pietrinferni}, {Cassisi}, {Salaris},
  {Percival}, \& {Ferguson}}]{Pietrinferni09}
{Pietrinferni}, A., {Cassisi}, S., {Salaris}, M., {Percival}, S., \&
  {Ferguson}, J.~W. 2009, \apj, 697, 275

\bibitem[{{Piotto}(2008)}]{Piotto08}
{Piotto}, G. 2008, \memsai, 79, 334

\bibitem[{{Piotto}(2009)}]{Piotto09}
{Piotto}, G. 2009, in IAU Symposium, Vol. 258, IAU Symposium, ed. E.~E.
  {Mamajek}, D.~R. {Soderblom}, \& R.~F.~G. {Wyse}, 233--244

\bibitem[{{Piotto} {et~al.}(2007){Piotto}, {Bedin}, {Anderson}, {King},
  {Cassisi}, {Milone}, {Villanova}, {Pietrinferni}, \& {Renzini}}]{Piotto07}
{Piotto}, G., {Bedin}, L.~R., {Anderson}, J., {et~al.} 2007, \apjl, 661, L53

\bibitem[{{Planck Collaboration} {et~al.}(2013){Planck Collaboration}, {Ade},
  {Aghanim}, {Armitage-Caplan}, {Arnaud}, {Ashdown}, {Atrio-Barandela},
  {Aumont}, {Baccigalupi}, {Banday}, \& et~al.}]{Planck13}
{Planck Collaboration}, {Ade}, P.~A.~R., {Aghanim}, N., {et~al.} 2013, ArXiv
  e-prints

\bibitem[{{Prantzos} \& {Charbonnel}(2006)}]{Prantzos06}
{Prantzos}, N. \& {Charbonnel}, C. 2006, \aap, 458, 135

\bibitem[{{Prantzos} {et~al.}(2007){Prantzos}, {Charbonnel}, \&
  {Iliadis}}]{Prantzos07}
{Prantzos}, N., {Charbonnel}, C., \& {Iliadis}, C. 2007, \aap, 470, 179

\bibitem[{{Ram{\'{\i}}rez} {et~al.}(2012){Ram{\'{\i}}rez}, {Mel{\'e}ndez}, \&
  {Chanam{\'e}}}]{Ramirez12}
{Ram{\'{\i}}rez}, I., {Mel{\'e}ndez}, J., \& {Chanam{\'e}}, J. 2012, \apj, 757,
  164

\bibitem[{{Reimers}(1975)}]{Reimers75}
{Reimers}, D. 1975, Memoires of the Societe Royale des Sciences de Liege, 8,
  369

\bibitem[{{Salaris} \& {Cassisi}(2005)}]{Salaris05}
{Salaris}, M. \& {Cassisi}, S. 2005, {Evolution of Stars and Stellar
  Populations}

\bibitem[{{Salaris} \& {Cassisi}(2014)}]{SalarisCassisi14}
{Salaris}, M. \& {Cassisi}, S. 2014, \aap, 566, A109

\bibitem[{{Sbordone} {et~al.}(2011){Sbordone}, {Salaris}, {Weiss}, \&
  {Cassisi}}]{Sbordone11}
{Sbordone}, L., {Salaris}, M., {Weiss}, A., \& {Cassisi}, S. 2011, \aap, 534,
  A9

\bibitem[{{Siess}(2007)}]{Siess07}
{Siess}, L. 2007, \aap, 476, 893

\bibitem[{{Sills} \& {Glebbeek}(2010)}]{Sills10}
{Sills}, A. \& {Glebbeek}, E. 2010, \mnras, 407, 277

\bibitem[{{Strickler} {et~al.}(2009){Strickler}, {Cool}, {Anderson}, {Cohn},
  {Lugger}, \& {Serenelli}}]{Strickler09}
{Strickler}, R.~R., {Cool}, A.~M., {Anderson}, J., {et~al.} 2009, \apj, 699, 40

\bibitem[{{Sweigart}(1978)}]{Sweigart78}
{Sweigart}, A.~V. 1978, in IAU Symposium, Vol.~80, The HR Diagram - The 100th
  Anniversary of Henry Norris Russell, ed. A.~G.~D. {Philip} \& D.~S. {Hayes},
  333--343

\bibitem[{{Valcarce} {et~al.}(2012){Valcarce}, {Catelan}, \&
  {Sweigart}}]{Valcarce12}
{Valcarce}, A.~A.~R., {Catelan}, M., \& {Sweigart}, A.~V. 2012, \aap, 547, A5

\bibitem[{{Vassiliadis} \& {Wood}(1993)}]{Vassiliadis93}
{Vassiliadis}, E. \& {Wood}, P.~R. 1993, \apj, 413, 641

\bibitem[{{Ventura} \& {D'Antona}(2011)}]{Ventura11}
{Ventura}, P. \& {D'Antona}, F. 2011, \mnras, 410, 2760

\bibitem[{{Ventura} {et~al.}(2001){Ventura}, {D'Antona}, {Mazzitelli}, \&
  {Gratton}}]{Ventura01}
{Ventura}, P., {D'Antona}, F., {Mazzitelli}, I., \& {Gratton}, R. 2001, \apjl,
  550, L65

\bibitem[{{Ventura} {et~al.}(2013){Ventura}, {Di Criscienzo}, {Carini}, \&
  {D'Antona}}]{Ventura13}
{Ventura}, P., {Di Criscienzo}, M., {Carini}, R., \& {D'Antona}, F. 2013,
  \mnras, 431, 3642

\bibitem[{{Villanova} {et~al.}(2009){Villanova}, {Piotto}, \&
  {Gratton}}]{Villanova09}
{Villanova}, S., {Piotto}, G., \& {Gratton}, R.~G. 2009, \aap, 499, 755

\bibitem[{{Villanova} {et~al.}(2007){Villanova}, {Piotto}, {King}, {Anderson},
  {Bedin}, {Gratton}, {Cassisi}, {Momany}, {Bellini}, {Cool}, {Recio-Blanco},
  \& {Renzini}}]{Villanova07}
{Villanova}, S., {Piotto}, G., {King}, I.~R., {et~al.} 2007, \apj, 663, 296

\bibitem[{{Webbink}(1975)}]{Webbink75}
{Webbink}, R.~F. 1975, \mnras, 171, 555

\end{thebibliography}

\end{document}